\documentclass[12pt]{article}

\usepackage{amsmath,amssymb}
\usepackage{slashed}
\usepackage{xcolor} 
\usepackage{graphicx}
\usepackage{url}
\usepackage{cite}
\usepackage{pdflscape}

\def\lsim{\mathrel{\rlap{\lower4pt\hbox{\hskip1pt$\sim$}}
  \raise1pt\hbox{$<$}}}
\def\gsim{\mathrel{\rlap{\lower4pt\hbox{\hskip1pt$\sim$}}
  \raise1pt\hbox{$>$}}}

\newcommand{\ba}{\begin{array}}
\newcommand{\ea}{\end{array}}
\newcommand{\beq}{\begin{equation}}
\newcommand{\eeq}{\end{equation}}
\newcommand{\bea}{\begin{eqnarray}}
\newcommand{\eea}{\end{eqnarray}}
\newcommand{\bvec}{\left ( \ba{c}}
\newcommand{\evec}{\ea \right )}
\newcommand{\bi}{\begin{itemize}}
\newcommand{\ei}{\end{itemize}}

\newcommand{\nnl}{\nonumber \\ }

\newcommand{\hc}{\mathrm{h.c.}}

\newcommand{\cO}{{\cal O}}
\newcommand{\cL}{{\cal L}}

\graphicspath{{./Figures/}}

\newcommand{\eref}[1]{Eq.~\eqref{eq:#1}}
\newcommand{\aref}[1]{Appendix~\ref{app:#1}}
\newcommand{\sref}[1]{Section~\ref{sec:#1}}
\newcommand{\tref}[1]{Table~\ref{tab:#1}}

\newcommand{\im}{\mathbb{I}}

\newenvironment{Eqnarray}{\arraycolsep 0.14em\begin{eqnarray}}{\end{eqnarray}}
\def\beqa{\begin{Eqnarray}}
\def\eeqa{\end{Eqnarray}}

\begin{document}


\begin{center}
\vspace*{15mm}

\vspace{1cm}
{\large \bf
Electroweak constraints on flavorful effective theories
} \\
\vspace{1.4cm}

{Aielet Efrati$\,^{a}$, Adam Falkowski$\,^{b}$,  and Yotam Soreq$\,^{a}$.}

 \vspace*{.5cm}
$^a$Department of Particle Physics and Astrophysics, Weizmann Institute of Science, Rehovot 7610001, Israel. \\
$^b$Laboratoire de Physique Th\'{e}orique, Bat.~210, Universit\'{e} Paris-Sud, 91405 Orsay, France.

\vspace*{.2cm}

\end{center}

\vspace*{10mm}
\begin{abstract}\noindent\normalsize
We derive model-independent constraints arising from the $Z$ and $W$ boson observables on dimension six operators in the effective theory beyond the Standard Model. In particular, we discuss the generic flavor structure for these operators as well as several flavor patterns motivated by simple new physics scenarios.
\end{abstract}

\vspace*{3mm}
\newpage
\renewcommand{\theequation}{\arabic{section}.\arabic{equation}}

\section{Introduction}
\setcounter{equation}{0}

If new particles beyond the Standard Model~(SM) are much heavier than the weak scale, their effects on current collider experiments can be described without introducing new degrees of freedom. This can be done in the framework of an effective field theory~(EFT) with the SM Lagrangian supplemented by higher-dimensional operators constructed out of only SM fields~\cite{Buchmuller:1985jz,Grzadkowski:2010es}. The EFT Lagrangian is organized as an expansion in operator dimensions $D$. The SM Lagrangian, which contains the renormalizable operators, is the leading term in this expansion. Assuming lepton number conservation,  the next-to-leading contributions to physical observables come from dimension six operators, $\mathcal{O}_6$.

In the upcoming years, the LHC and other experiments will be searching for multiple signatures of~$\mathcal{O}_6$. From this perspective, it is important  to understand what are the existing constraints on these operators from previous measurements. In particular, one should assess the constraining power of electroweak~(EW) precision measurements with on-shell $Z$ or $W$ bosons, which are among the most accurately measured observables in collider physics. Such studies have a long history, see for example  \cite{Han:2004az,Han:2005pr,Barbieri:2004qk,Grojean:2006nn,Cacciapaglia:2006pk,Pomarol:2013zra,Elias-Miro:2013mua,Dumont:2013wma,Chen:2013kfa,Willenbrock:2014bja,Gupta:2014rxa,Masso:2014xra,deBlas:2014ula,Ciuchini:2014dea,Ellis:2014huj,Falkowski:2014tna,Berthier:2015oma}. However, constraints in the general situation where all $D=6$ operators can be  simultaneously present have not been derived so far. In particular, previous analyses typically assumed that the coefficients of the dimension six operators involving the SM fermions do not depend on the fermion generation index or assume a non-generic flavor structure~\cite{Han:2005pr}. This is justified by the humongous number of $\mathcal{O}_6$ once a general flavor structure is allowed~\cite{Alonso:2013hga}. However, this situation is not completely satisfactory, since many well motivated scenarios predict $D=6$ operators in the low-energy EFT that are not flavor universal. It is important to determine whether the strong bounds on these operators obtained under the assumption of flavor universality~\cite{Pomarol:2013zra,Ciuchini:2014dea,Ellis:2014huj,Falkowski:2014tna} are robust and survive in  a completely generic scenario. Moreover, understanding of the weakest constrained directions in the flavor space is important both for model building and to identify the promising experimental signatures.

In this paper we consider an EFT where the higher-dimensional operator have a completely arbitrary flavor structure.
In such a setting, we derive constraints on a subset of $\mathcal{O}_6$ that affect the $W$ boson mass and the $Z$ or $W$ boson couplings to fermions. Our constraints are based on the {\em pole observables} where a single $Z$ or $W$ boson is produced {\em on-shell}.
Contributions of 4-fermion operators to these processes are suppressed by the $Z$ or $W$ boson width over its mass, as compared to contributions of the 2-fermion operators, which is roughly an $\cO(3\%)$ correction~\cite{Han:2004az}. We therefore neglect all 4-fermion operators (apart from one that contributes to our input parameters) in our analysis, reducing the number of operators to a tractable set. All in all, the pole observables depend only on those $D=6$ operators that modify the $Z$ and $W$ couplings to fermions (so-called {\em vertex corrections}) or electroweak gauge boson propagators (so-called {\em oblique corrections}), or affect the relation between electroweak parameters and input observables.

To calculate the corrections to physical observables one needs to choose a basis of $\mathcal{O}_6$. In this paper we use the basis advertised in Refs.~\cite{Gupta:2014rxa,Pomarol:2014dya,HXSWGbasis}.
Rather than parameterizing observables in terms of Wilson coefficients of $SU(3) \times SU(2) \times U(1)$ invariant operators, we use to this end the couplings of SM mass eigenstates after electroweak symmetry breaking.
The $SU(2) \times U(1)$ symmetry of $D=6$ operators is not manifest in this language;  instead it is  encoded in the relations between different  couplings in the mass eigenstate Lagrangian.
This formalism is particularly convenient to connect the EFT to collider observables.
In this approach, all oblique corrections are redefined away, with the exception of  the correction to the $W$  boson mass.
Once that is achieved, the only parameters affecting the pole observables  at the leading order are the vertex corrections $\delta g$ and the $W$ mass correction $\delta m$.
This way, the relevant parameters for the pole observables are clearly identified, without any unconstrained (flat) combination of parameters among them.
To translate these constraints to another basis, $\delta g$ and $\delta m$ should be mapped to a linear combinations of $D=6$ operators in that basis.
We provide such a mapping for one particular basis reviewed in \aref{warsaw}.

Our work shows that the existing measurements of the pole observables simultaneously constrain $\delta m$ and 20 independent vertex corrections to flavor-diagonal $W$ and $Z$ interactions in the SM  (only vertex corrections to the $Z t_R t_R$ coupling cannot be constrained by our analysis). Some off-diagonal vertex corrections to $Z$ boson couplings to quarks and leptons can also be constrained. The strength of the limits varies depending on the interaction in question. For example, the corrections to the $W$ boson mass and the leptonic couplings of $Z$ are most strongly constrained, at the level of $\cO(10^{-4})-\cO(10^{-3})$. On the other hand, couplings of the first generation and the right-handed top quarks are only weakly constrained by current data, at the level of $\cO(10^{-1})-\cO(1)$. Relying on CP even observables, our analysis has no sensitivity to complex phases in the Wilson coefficients of dimension six operators.

Specific flavor models predict different correlations between the various vertex corrections. For instance, if the UV theory is flavor universal it will induce flavor universal vertex corrections with no flavor changing neutral currents~(FCNC). However, any deviation from universality will, in general, lead to FCNC due to the misalignment of the up- and down-type left-handed quarks. In this work we consider three flavor scenarios which address the new physics (NP) flavor puzzle: alignment~\cite{Nir:1993mx,Leurer:1993gy}, Minimal Flavor Violation~(MFV)~\cite{D'Ambrosio:2002ex,Hall:1990ac,Chivukula:1987py,Buras:2000dm} and anarchic  partial compositeness or warped extra dimensions~\cite{ArkaniHamed:1999dc, Gherghetta:2000qt, Huber:2000ie,Agashe:2004cp,Agashe:2008fe,Contino:2006nn,KerenZur:2012fr} which is similar to vector-like fermion scenario~\cite{Falkowski:2013jya,Blum:2015rpa}. 
In the first two class of models the magnitude of the off-diagonal couplings is dictated by the non-universality of the diagonal vertex corrections, resulting indirectly in stringent constraints on the off-diagonal couplings. An analysis of $B$ and top FCNC in alignment EFT based on a covariant description is given in Ref.~\cite{Gedalia:2010zs,Gedalia:2010mf} and the universality of CP violation in $\Delta F=1$ processes is pointed out~\cite{Gedalia:2012pi}. 

This paper is organized as follows.~\sref{effl} introduces our formalism and notation. In~\sref{cons} we present the experimental data and theoretical premises used in our work.~\sref{res} (and~\aref{rho21}) contains our results in the completely generic  case, as well as within the above mentioned flavor scenarios. We conclude in~\sref{con}.~\aref{otherbases} details the relations between the parameters  we constrain in our formalism and the Wilson coefficients of the dimension six operators in two  particular bases - the Warsaw basis proposed in Ref.~\cite{Grzadkowski:2010es} and the SILH basis proposed in Ref.~\cite{Giudice:2007fh}. For completeness, we analyze the constraints on the off-diagonal couplings to quarks arising from low-energy observables in~\aref{indirectZoff}.

\section{Preliminaries}
\setcounter{equation}{0}
\label{sec:effl}

We start by briefly summarizing our conventions and notations. The $SU(3) \times SU(2) \times U(1)$ gauge couplings of the SM are denoted by $g_s$, $g_L$, $g_Y$; we also define the electromagnetic coupling $e = g_L s_\theta$, where $s_\theta = g_Y/\sqrt{g_L^2 + g_Y^2}$ is the Weinberg angle. The Higgs doublet $(H)$ acquires a Vacuum Expectation Value~(VEV): $\langle  H^\dagger H \rangle = v^2/2$, spontaneously breaking EW symmetry. For the SM fermions we employ the two-component spinor notation, with all conventions inherited from Ref.~\cite{Dreiner:2008tw}.
The left-handed spinors of the up-type quarks, down-type quarks, and charged leptons are denoted by $u,u^c$, $d,d^c$, $e,e^c$, and neutrinos are denoted as $\nu$.
All fermions are three vectors in generation space. We work in the mass eigenstate basis in which $\cL_{m} = - \sum_{f_i} m_{f_i}  f_i f_i^c + \hc$ where $m$ is diagonal.

We consider the effective Lagrangian of the form,
\beq
\label{eq:leff}
\cL_{\rm eff}  = {\cal L}^ {\rm SM}  +  {1 \over v^2} {\cal L}^ {D=6}, \qquad   {\cal L}^ {D=6} = \sum_i c_i \mathcal{O}_{6,i},
\eeq
where $\cL^{\rm SM}$ is the SM Lagrangian,  while $\mathcal{O}_{6,i}$ is a complete basis of $SU(3) \times SU(2) \times U(1)$ invariant $D=6$ operators constructed out of the SM fields. Any such basis contains 2499 independent operators after imposing baryon and lepton number conservation~\cite{Alonso:2013hga}. However, working at tree level, a much smaller subset is relevant for electroweak precision observables. The couplings in  the effective Lagrangian are defined at the scale  $m_Z$; we neglect running and mixing effect to other relevant scales as $m_W$ or $m_t$ which are subleading in our analysis, a detailed discussion on these effects can be found in~\cite{Jenkins:2013zja}.

As mentioned previously, we parameterize the effect of $\mathcal{O}_6$ on the interactions of the SM mass eigenstates, rather than writing down a specific basis of $D=6$ operators. 
We work with an effective Lagrangian where all mass terms and kinetic terms are diagonal, using the $Z$ boson mass as an input parameter (hence introducing no correction to the $Z$ mass term). While, in general, $D=6$ operators do generate such mixing and mass corrections, the canonical form can always be recovered by using the equations of motion, integration by parts, and redefinition of the fields and  couplings. In this basis, the gauge boson mass terms take the form
\beq
\label{eq:lvv}
\cL_{\rm eff}^{vv} = {(g_L^2 + g_Y^2) v^2 \over 8} Z_\mu Z_\mu +  {g_L^2 v^2 \over 4} \left (1 + 2 \delta m \right) W_\mu^+ W_\mu^- ,
 \eeq
where $\delta m$ parameterizes the corrections to the $W$ boson mass from $D=6$ operators.
The interactions between the SM gauge bosons and fermions are then
\bea
\label{eq:lvff}
\cL_{\rm eff}^{v ff} & =  &
 e A_\mu \sum_{f  \in u,d,e} Q_f (\bar f \bar \sigma_\mu f  + f^c \sigma_\mu  \bar f^c)   + g_s G_\mu^a \sum_{f  \in u,d} (\bar f \bar \sigma_\mu T^a f  + f^c \sigma_\mu T^a \bar f^c)
\\ &+&
{g_L \over \sqrt 2}  \left (
W_\mu^+ \bar u \bar \sigma_\mu ( V+  \delta g^{Wq}_L )  d
+
W_\mu^+  \bar u \bar \sigma_\mu \delta g^{Wq}_R  d_R
+  W_\mu^+  \bar \nu \bar \sigma_\mu (\im +  \delta g^{W\ell}_L  ) e+  \hc \right )
\nnl &+& \sqrt{g_L^2 + g_Y^2} Z_\mu      \left [
   \sum_{f \in u, d,e,\nu} \bar f \bar \sigma_\mu ( \im  T^3_f  -  \im  s^2_\theta Q_f + \delta g^{Zf}_L) f
+   \sum_{f^c \in u^c, d^c,e^c} f^c \sigma_\mu (- \im  s^2_\theta Q_f  + \delta g^{Zf}_R)\bar  f^c  \right ],
\nonumber
\eea
where $ \im$ is the $3\times 3$ unit matrix, and $V$ is the CKM matrix. 
The effects of $D=6$ operators are parameterized by the vertex corrections $\delta g$,
which are $3 \times 3$ matrices in the generation space with, in general, non-diagonal elements.
The local  $SU(2) \times U(1)$ symmetry of the effective Lagrangian implies the following relations:
\beq
\label{eq:lvff2}
 \delta g^{Z\nu}_L  =   \delta g^{Ze}_L   +  \delta g^{W \ell}_L , \qquad
\delta g^{Wq}_L =  \delta g^{Zu}_L V-   V  \delta g^{Zd}_L .
\eeq
Note that the gauge interactions of the photon and the gluon in~\eref{lvff} are  the same as in the SM; again, this can be always ensured without loss of generality via  redefinitions of fields and couplings. The relation between the vertex corrections and the Wilson coefficients in the basis  of Ref.~\cite{Grzadkowski:2010es} is given in \aref{warsaw}.

To summarize, the effects of $\mathcal{O}_6$ relevant for EW pole observables is parameterized using
\beq
\label{eq:primary_ewpt}
 \delta m, \ \delta g^{W \ell}_L,  \  \delta g^{Ze}_L, \ \delta g^{Ze}_R, \ \delta g^{Zu}_L,  \ \delta g^{Zu}_R,   \  \delta g^{Zd}_L,  \ \delta g^{Zd}_R,  \ \delta g^{Wq}_R, 
\eeq
which stand for $1+7\times6+9=52$ real parameters in the general case (plus 30 complex phases which we are not sensitive to in this analysis.)

\section{Electroweak Observables}
\setcounter{equation}{0}
\label{sec:cons}

In this section we list the experimental data we use and the corresponding SM predictions for the pole observables. We further draw our assumptions and the statistical treatment we take. The relevant observables are summarized in~\tref{EWPT_zpole},~\tref{EWPT_wpole} and~\tref{OffDiag}. Starting with~\eref{lvv} and~\eref{lvff}, we calculate the leading corrections to these observables in terms of the effective Lagrangian parameters $\delta m$ and $\delta g$, and the SM input parameters $g_L$, $ g_Y$ and $v$.

The basic premises of our procedure are the following:
\bi
\item
For the SM predictions of the pole observables, we use the state-of-art  theoretical calculations.
Whenever available, we use the central value quoted in Table~2 of Ref.~\cite{Baak:2014ora}.
We ignore  the theoretical errors, which are subleading compared to the experimental ones.
We verified that including the theoretical errors does not affect our results in an appreciable way.
\item The electroweak parameters (that we need to evaluate NP corrections) are extracted at tree-level from the muon lifetime $\tau_\mu =  384 \pi^3 v^4/m_\mu^5$  (equivalently, from the Fermi constant $G_F = 1/\sqrt 2 v^2 = 1.16637 \times 10^{-5}~{\rm GeV}^{-2}$~\cite{Beringer:1900zz}),  the  electromagnetic constant $\alpha(m_Z) = e^2/4\pi = 7.755 \times 10^{-3}$ \cite{Burkhardt:2011ur},  and the $Z$ boson mass $m_Z = \sqrt{g_L^2 + g_Y^2} v/2= 91.1875$~GeV \cite{ALEPH:2005ab}.
With this choice, the tree-level values of the electroweak parameters are
\beq
g_L = 0.650,\quad\quad\quad g_Y =  0.356,\quad\quad\quad v = 246.2{\rm ~GeV}.
\eeq
\item
We work at the level of $D=6$ operators neglecting possible contributions of dimension-8 operators.
Consistently, for observables where the SM contribution is non-zero, we only include the leading corrections that are formally $\cO(v^2/\Lambda^2)$ in the EFT counting.
These come from interference terms between NP and SM contributions to the amplitudes of the relevant processes, and they are {\em linear} in $\delta m$ and $\delta g$.
Quadratic corrections in $\delta g$ and $\delta m$ are in this case neglected, since they are formally of order $\cO(v^4/\Lambda^4)$, much as the contributions from dimension-8 operators that we ignore.
\item
The off-diagonal neutral current couplings are absent in the SM at the tree level. The leading order contribution to the branching ratios for flavor violating $Z$ decays is therefore $\cO(v^4/\Lambda^4)$, and {\em quadratic} in $\delta g$. In this case, the contribution from  possible dimension-8 operators is  parametrically $\cO(v^6/\Lambda^6)$,  and, again, can be neglected. 
Similarly, the effects of flavor-diagonal vertex corrections on flavor-violating $Z$ decays (that enter via corrections to the total $Z$ width) are parametrically $\cO(v^6/\Lambda^6)$ and are neglected.
\item
We ignore all loop-suppressed effects proportional to $\delta g$ and $\delta m$.
In particular, we only take into account the interference terms between tree-level NP corrections and tree-level SM contributions, while we ignore the interference of the NP corrections with loop-level SM contributions.
This is the largest source of uncertainty on the central values and standard deviations of $\delta g$ and $\delta m$ that we quote below.
From the change of the limits under variation of the input electromagnetic coupling between the scale $m_Z/2$ and $2m_Z$ we estimate this uncertainty to be of order 30\%.
\item
All the observables we consider are measured for $Z$ or $W$ bosons close to the mass shell. Thanks to that, we can ignore the contribution of 4-fermion operators, which is suppressed by $\Gamma_Z/m_Z$ or $\Gamma_W/m_W$ \cite{Han:2004az,Berthier:2015oma}.
The only exception is the  $V_{tb}$ measurement extracted from the single $t$-channel top production at the LHC; in this case,  the experimental cuts suppress possible contributions of 4-fermion operators to this observable. 
\item
We neglect CKM-suppressed corrections. As a result, the pole observables depend only on the diagonal elements of $\delta g$.
Furthermore, corrections proportional to $\delta g^{Wq}_R$ do not interfere with the SM amplitudes; therefore they enter only quadratically and are neglected.
\ei
All in all, at the tree level, the pole observables depend linearly on $3 \times 7 - 1= 20$ diagonal elements of $\delta g^{Ze}_L$, $\delta g^{Ze}_R$, $\delta g^{W \ell}_L$,  $\delta g^{Zu}_L$,  $\delta g^{Zu}_R$,  $\delta g^{Zd}_L$, $\delta g^{Zd}_R$ and on $\delta m$ (they do not depend on the $Z$ coupling to right-handed top quarks). All these couplings are simultaneously constrained by the observables $O_{i}$ listed in~\tref{EWPT_zpole} and ~\tref{EWPT_wpole}. Moreover, 4 combinations of the $Z$ off-diagonal couplings are constrained by the limits listed in~\tref{OffDiag}.

To construct a global $\chi^2$ function, we write the observables as
\beq
O_{i,\rm th}  = O_{i,\rm SM}^{\rm NNLO}  +  \vec{\delta g} \cdot \vec O_{i,\rm BSM}^{\rm LO}
\eeq
The state-of-art SM predictions $O_{i,\rm SM}^{\rm NNLO}$ are provided in the literature,
while the tree-level NP corrections $\vec{\delta g}\cdot \vec O_{i,\rm BSM}^{\rm LO}$  linear  in $\delta g$ are computed analytically.
Then $\chi^2$ function is constructed as
\beq
\chi^2 = \sum_{ij}\left [ O_{i,\rm exp}  -  O_{i,\rm th} \right ] \sigma^{-2}_{ij}  \left [ O_{j,\rm exp}  -  O_{j,\rm th} \right ],
\eeq
where  $\sigma^{-2}_{ij} =  [\delta O_{i} \rho_{ij, {\rm exp}} \delta O_{j} ]^{-1}$ is calculated from the known experimental errors $\delta O_{i}$ and their correlations $\rho_{ij, \rm exp}$ (whenever quoted).

 \begin{table}
 \begin{center}
 \begin{tabular}{|c|c|c|c|c|}
\hline
{\color{blue}{Observable}} & {\color{blue}{Experimental value}}   &   {\color{blue}{Ref.}}   &  {\color{blue}{SM prediction}}    &  {\color{blue} Definition}
  \\  \hline   \hline
$\Gamma_{Z}$ [GeV]  & $2.4952 \pm 0.0023$ & \cite{ALEPH:2005ab} & $ 2.4950$    & $\sum_f \Gamma (Z \to f \bar f)$
 \\  \hline
$\sigma_{\rm had}$ [nb]  & $41.541\pm 0.037$ &\cite{ALEPH:2005ab} &  $41.484$ &  ${12 \pi \over m_Z^2} {\Gamma (Z \to e^+ e^-) \Gamma (Z \to q \bar q) \over \Gamma_Z^2}$
  \\  \hline
 $R_{e}$  & $20.804\pm 0.050$ & \cite{ALEPH:2005ab} &  $20.743$  &    $ {\sum_{q} \Gamma(Z \to q \bar q) \over  \Gamma(Z \to e^+ e^-)} $
   \\  \hline
 $R_{\mu}$  & $20.785 \pm 0.033$ & \cite{ALEPH:2005ab} &  $20.743$  &    $ {\sum_{q} \Gamma(Z \to q \bar q) \over  \Gamma(Z \to \mu^+ \mu^-)} $
   \\  \hline
 $R_{\tau}$  & $20.764\pm 0.045$ & \cite{ALEPH:2005ab} &  $20.743$  &    $ {\sum_{q} \Gamma(Z \to q \bar q) \over  \Gamma(Z \to \tau^+ \tau^-)} $
 \\  \hline
 $A_{\rm FB}^{0,e}$ & $0.0145\pm 0.0025$ &\cite{ALEPH:2005ab} &  $0.0163$   &  ${3 \over 4} A_e^2$
  \\  \hline
 $A_{\rm FB}^{0,\mu}$ & $0.0169\pm 0.0013$ &\cite{ALEPH:2005ab} &  $0.0163$   &  ${3 \over 4} A_e A_\mu $
  \\  \hline
 $A_{\rm FB}^{0,\tau}$ & $0.0188\pm 0.0017$ &\cite{ALEPH:2005ab} &  $0.0163$   &  ${3 \over 4} A_e A_\tau$
  \\ \hline \hline
$R_b$ & $0.21629\pm0.00066$ & \cite{ALEPH:2005ab} & $0.21578$  &     ${ \Gamma(Z \to b \bar b) \over \sum_q \Gamma(Z \to q \bar q)}$
   \\  \hline
$R_c$ & $0.1721\pm0.0030$  & \cite{ALEPH:2005ab}  & $0.17226$   & ${ \Gamma(Z \to c \bar c) \over \sum_q \Gamma(Z \to q \bar q)} $
\\  \hline
$A_{b}^{\rm FB}$ & $0.0992\pm 0.0016$ & \cite{ALEPH:2005ab}  & $0.1032$  & ${3 \over 4} A_e A_b$
 \\  \hline
 $A_{c}^{\rm FB}$ & $0.0707\pm 0.0035$  & \cite{ALEPH:2005ab} &  $0.0738$  & ${3 \over 4} A_e A_c$
  \\ \hline \hline
 $A_e$ & $0.1516 \pm 0.0021$ &\cite{ALEPH:2005ab} &  $0.1472$ &  ${\Gamma(Z \to e_L^+ e_L^-) - \Gamma(Z \to e_R^+ e_R^-) \over \Gamma(Z \to e^+  e^-) }$
 \\ \hline
  $A_\mu$ & $0.142 \pm 0.015$ &\cite{ALEPH:2005ab} &  $0.1472$ &  ${\Gamma(Z \to \mu_L^+ \mu_L^-) - \Gamma(Z \to \mu_R^+ \mu_R^-) \over \Gamma(Z \to \mu^+  \mu^-) }$
 \\ \hline
 $A_\tau$ & $0.136 \pm 0.015$ &\cite{ALEPH:2005ab} &  $0.1472$ &  ${\Gamma(Z \to \tau_L^+ \tau_L^-) - \Gamma(Z \to \tau_R^+ \tau_R^-) \over \Gamma(Z \to \tau^+  \tau^-) }$
 \\ \hline \hline
  $A_e$ & $0.1498 \pm 0.0049$ & \cite{ALEPH:2005ab} &  $0.1472$ &  ${\Gamma(Z \to e_L^+ e_L^-) - \Gamma(Z \to e_R^+ e_R^-) \over \Gamma(Z \to \tau^+  \tau^-) }$
 \\ \hline \hline
 $A_\tau$ & $0.1439 \pm  0.0043$ & \cite{ALEPH:2005ab} &  $0.1472$ &  ${\Gamma(Z \to \tau_L^+ \tau_L^-) - \Gamma(Z \to \tau_R^+ \tau_R^-) \over \Gamma(Z \to \tau^+  \tau^-) }$
 \\ \hline \hline
 $A_b$ & $0.923\pm 0.020$ & \cite{ALEPH:2005ab} & $0.935$  &
 ${ \Gamma(Z \to b_L \bar b_L)  -   \Gamma(Z \to b_R \bar b_R)  \over  \Gamma(Z \to b \bar b)   }$
 \\  \hline
$A_c$ & $0.670 \pm 0.027$ & \cite{ALEPH:2005ab} & $0.668$
 &  ${ \Gamma(Z \to c_L \bar c_L)  -   \Gamma(Z \to c_R \bar c_R)  \over  \Gamma(Z \to c \bar c)   }$
 \\ \hline \hline
$A_s$ & $0.895 \pm 0.091$ & \cite{Abe:2000uc} & $0.935$
&  ${ \Gamma(Z \to s_L \bar s_L)  -   \Gamma(Z \to s_R \bar s_R)  \over  \Gamma(Z \to s\bar s)   }$
\\ \hline \hline
$R_{uc}$ & $0.166 \pm 0.009$  & \cite{Beringer:1900zz}  & $0.1724$   & ${ \Gamma(Z \to u \bar u) +  \Gamma(Z \to c \bar c) \over 2 \sum_q \Gamma(Z \to q \bar q)} $
\\ \hline \hline
\end{tabular}
\end{center}
\caption{
{\bf $Z$ boson pole observables}. The experimental errors of the observables between the double lines are correlated, which is taken into account in the fit. $A_e$ and $A_\tau$  are listed twice:  the first number  comes from the combination of leptonic polarization and left-right asymmetry measurements at the SLC collider, while the second from the tau polarization measurements at LEP-1. We also include the model-independent measurement of on-shell $Z$ boson couplings to light quarks in D0~\cite{Abazov:2011ws}. For the theoretical predictions we use the best fit SM values from GFitter~\cite{Baak:2014ora}.
}
\label{tab:EWPT_zpole}
 \end{table}

\begin{table}
 \begin{center}
 \begin{tabular}{|c|c|c|c|c|}
 \hline
{\color{blue}{Observable}} & {\color{blue}{Experimental value}}   &   {\color{blue}{Ref.}}   &  {\color{blue}{SM prediction}}    &  {\color{blue} Definition}
 \\  \hline  \hline
 $m_{W}$ [GeV]  & $80.385 \pm 0.015$ &\cite{Group:2012gb}    &  $80.364$   &  ${g_L v \over 2} \left ( 1 + \delta m \right )$
\\ \hline  \hline
$\Gamma_{W}$ [GeV]  & $ 2.085 \pm 0.042$  & \cite{Beringer:1900zz} &  $2.091$  &  $ \sum_f  \Gamma(W \to f f')$
\\ \hline \hline
${\rm Br} (W \to e \nu)$ & $ 0.1071 \pm 0.0016$ &\cite{Schael:2013ita} &  $0.1083$  &  $ { \Gamma(W \to e \nu) \over  \sum_f  \Gamma(W \to f f')}$
\\ \hline
${\rm Br} (W \to \mu \nu)$ & $ 0.1063 \pm 0.0015$ &\cite{Schael:2013ita} &  $0.1083$  &  $ { \Gamma(W \to \mu \nu) \over  \sum_f  \Gamma(W \to f f')}$
\\ \hline
${\rm Br} (W \to \tau \nu)$ & $ 0.1138 \pm 0.0021$ &\cite{Schael:2013ita} &  $0.1083$  &  $ {\Gamma(W \to \tau \nu) \over  \sum_f  \Gamma(W \to f f')}$
 \\ \hline \hline
 $R_{Wc}$ & $ 0.49 \pm 0.04$ & \cite{Beringer:1900zz}  &  $0.50$  &  $ { \Gamma(W \to c s) \over  \Gamma(W \to u d) + \Gamma(W \to c s) }$
 \\ \hline \hline
 $R_{\sigma}$ & $0.998 \pm 0.041$  & \cite{Khachatryan:2014iya} & 1.000 & $g^{Wq_3}_L/g^{Wq_3}_{L,\rm SM} $
  \\ \hline \hline
\end{tabular}
\end{center}
\caption{{\bf $W$-boson pole observables}. 
We also include the $V_{tb}$ measurement in the single-top t-channel production at the LHC; even though W boson is not on-shell, the experimental cuts suppress possible contributions of 4-fermion operators to this observable. 
Measurements of the three leptonic branching ratios are correlated. For the theoretical predictions of $m_W$ and $\Gamma_W$, we use the best fit SM values from GFitter~\cite{Baak:2014ora}, while for the leptonic branching ratios we take the value quoted in  \cite{Schael:2013ita}.
}
\label{tab:EWPT_wpole}
 \end{table}
\begin{table}
 \begin{center}
 \begin{tabular}{|c|c|c|c|c|}
 \hline
{\color{blue}{Observable}} & {\color{blue}{Experimental bound}}   &   {\color{blue}{Ref.}}   &  {\color{blue} Definition}
 \\  \hline  \hline
 ${\rm Br} (Z \to e \mu)$  & $7.5 \times 10^{-7}$ &\cite{Aad:2014bca}   &  ${\Gamma(Z \to e \mu) \over  \sum_f  \Gamma(Z \to f f')}$
\\ \hline  \hline
 ${\rm Br} (Z \to e \tau)$  & $9.8 \times 10^{-6}$ &\cite{Akers:1995gz}   &  ${\Gamma(Z \to e \tau) \over  \sum_f  \Gamma(Z \to f f')}$
\\ \hline  \hline
 ${\rm Br} (Z \to \mu \tau)$  & $1.2 \times 10^{-5}$ &\cite{Abreu:1996mj}   &  ${\Gamma(Z \to \mu \tau) \over  \sum_f  \Gamma(Z \to f f')}$
\\ \hline  \hline
${\rm Br}(t \to $Z$ q)$  & $5.0 \times 10^{-4}$ &\cite{Chatrchyan:2013nwa}   &  ${\Gamma(t \to Z u)+\Gamma(t \to Z c) \over  \sum_f  \Gamma(Z \to f f')}$
\\ \hline  \hline
\end{tabular}
\end{center}
\caption{{\bf  Flavor-violating processes with $Z$-boson}. Limits are quoted at 95\% CL. 
In the SM, the lepton flavor violating Z decays completely vanish while the FCNC top decays are extremely suppressed to an unobservable level.
}
\label{tab:OffDiag}
 \end{table}

\section{Results}
\label{sec:res}

\subsection{Generic Scenario}

First, from the measurement of the $W$ boson mass we derive the constraint
\beq
\delta m = \left ( 2.6  \pm  1.9 \right ) \times 10^{-4}\, .
\eeq
The correlation between this result and the constraints on $\delta g$'s is small and will be neglected in the following.

Next, we derive the constraints on the $\delta g$'s  when all of them are simultaneously present and a-priori unrelated by the UV theory. 
Minimizing our $\chi^2$ function with respect to $\delta g$ we obtain the following central values and 1$\sigma$ errors:
\bea
\label{eq:dgl}
&&[\delta g^{We}_L]_{ii} = \bvec  -1.00 \pm 0.64 \\   -1.36  \pm 0.59 \\  1.95 \pm 0.79 \evec \times 10^{-2}, \\
\label{eq:dgZl}
&& [\delta g^{Ze}_L]_{ii} = \bvec  -0.26 \pm 0.28 \\   0.1  \pm 1.1 \\  0.16  \pm 0.58 \evec \times 10^{-3},
\quad [\delta g^{Ze}_R]_{ii} = \bvec  -0.37 \pm 0.27 \\   0.0  \pm 1.3 \\  0.39  \pm 0.62 \evec \times 10^{-3}, \\
\label{eq:dgZu}
&& [\delta  g^{Zu}_L]_{ii} = \bvec  -0.8 \pm 3.1 \\   -0.16  \pm 0.36 \\  -0.28  \pm 3.8 \evec \times 10^{-2},
\quad [ \delta g^{Zu}_R]_{ii} = \bvec   1.3 \pm 5.1 \\   -0.38  \pm 0.51  \\   \times  \evec \times 10^{-2}, \\
\label{eq:dgZd}
&& [ \delta  g^{Zd}_L]_{ii} = \bvec  -1.0 \pm 4.4 \\   0.9 \pm 2.8 \\   0.33  \pm 0.16 \evec \times 10^{-2},
\quad \quad [\delta  g^{Zd}_R]_{ii} = \bvec  2.9 \pm 16 \\ 3.5 \pm 5.0 \\  2.30 \pm 0.82 \evec \times 10^{-2}.
\eea
The corresponding $20 \times 20$ correlation matrix is given in \aref{rho21}.

As for the off diagonal couplings, we find:
\bea \label{eq:ZLFV}
\sqrt{|[\delta g_L^{Ze}]_{12}|^2 + | [\delta g_R^{Ze}]_{12}|^2} &<  & 1.2 \times 10^{-3},
\nnl
\sqrt{|[\delta g_L^{Ze}]_{13}|^2 + | [\delta g_R^{Ze}]_{13}|^2} &<  & 4.3 \times 10^{-3},
\nnl
\sqrt{|[\delta g_L^{Ze}]_{23}|^2 + | [\delta g_R^{Ze}]_{23}|^2} &<  & 4.8 \times 10^{-3},
\eea
where the measured central value of the $Z$ width is used and
\beq
\label{eq:tzc}
\sqrt{|[\delta g_L^{Zu}]_{13}|^2 + | [\delta g_R^{Zu}]_{13}|^2 + |[\delta g_L^{Zu}]_{23}|^2 + | [\delta g_R^{Zu}]_{23}|^2} <   1.6 \times 10^{-2}\left( \frac{\Gamma_t}{1.35\,{\rm GeV}}\right)^{1/2}\, ,
\eeq
at the 95\% CL. Here we take $\Gamma^{\rm SM}_t \simeq 1.35 \,$GeV for $m_t = 173$~GeV~\cite{Jezabek:1988iv}.

Using the above central values  $\delta g_0$, uncertainties $\delta g_\sigma$  and the correlation matrix $\rho$ one can reconstruct the dependence of the global  $\chi^2$ function on the vertex corrections:
\beq
\chi^2=   \sum_{ij}[\delta g - \delta g_0]_{i} \sigma^{-2}_{ij}   [\delta g - \delta g_0]_{j},
\eeq
where $\sigma^{-2}_{ij} =  [[\delta g_\sigma]_i \rho_{ij} [ \delta g_\sigma]_j]^{-1}$. In specific  extensions of the SM, the vertex corrections will be functions of a (typically smaller) number of  the model parameters. In this case,  the global $\chi^2$ function can be minimized with respect to the new parameters, and thus limits on this particular model can be obtained.
This way our results can be used to obtain the constraints on any specific UV model.

From our results for the vertex corrections,~\eref{dgl}--\eref{dgZd}, we learn the following:
\begin{itemize}
\item Globally, the fit is in a very good agreement with the SM, corresponding to the p-value of order~$40\%$.
\item Corrections to the $Z$ boson couplings to charged leptons are constrained at the level of $\cO(10^{-3})$. 
We stress that these stringent constraints are completely model-independent. On the other hand, $W$ couplings to leptons are somewhat less tightly constrained - at the level of $\cO(10^{-2})$ - than in the flavor universal case.
Due to the relation in~\eref{lvff2}, the $Z$ boson couplings to neutrinos are constrained with the same precision. 
\item As for the $Z$ boson couplings to quarks the situation is more complicated. Some of these couplings, specifically the ones to charm and bottom, are rather tightly constrained, at the level of $\cO(10^{-2})$. The couplings to top and strange quarks are weakly constrained, such that $\cO(10\%)$ deviations are possible, and the $Z$ coupling to the right-handed quarks is not constrained at all in a model-independent way. The couplings to the first generation quarks are poorly constrained in a model-independent way, especially the ones to right-handed down quarks.
\item In those cases where large couplings corrections are allowed, one needs to be more careful about the validity of the EFT expansion. The large corrections may be a result of large Wilson coefficients, where the higher dimensional operators can not be safely neglected. For example, when $\cO(30\%)$ corrections are allowed, this implies that higher dimensional operators suppressed by the scale $\sim 0.5$~TeV may be present in the Lagrangian. For new physics with order one couplings to the SM it  would  imply that the EFT expansion is inadequate (in particular, dimension-8 operators cannot be safely neglected). However, even in this case, the EFT expansion may be valid when new physics couples to the SM strongly, with the coupling close to the maximal value allowed by perturbativity.  
\item For some of the vertex corrections the best fit value is more than $2\sigma$ away from zero. 
In the case of $[\delta g^{Zd}_R]_{33} $ this reflects the famous anomaly in the forward-backward asymmetry of $b$-quark pair production at LEP-1; for  $[\delta g^{We}_L]_{33}$ it is due to the excess of the measured  $W\to\tau\nu$ at LEP-2.
\end{itemize}

One important comment is in order. The constraints on the vertex corrections we derived are valid in the Higgs basis, where oblique corrections are rotated away and new physics affects the pole observables via the effective Lagrangian in \eref{lvff}. Of course, physical observables are independent of a basis choice; however parametrization of new physics does depend on a basis. In another basis, larger parameters may be allowed
 if compensating oblique corrections are present~\cite{Grojean:2006nn,Cacciapaglia:2006pk}, such that physical corrections remain small.  For instance, in the Warsaw basis~\cite{Grzadkowski:2010es}, see \aref{warsaw} for details, both 2-fermions operators $O_{Hf}$ (that induce vertex corrections) and bosonic operators $O_{WB}$, $O_{T}$ (that induce oblique corrections) contribute to the pole observables. Neither the former nor the latter can be constrained by itself using the pole observables alone. In other words, there are 2 exactly flat directions of the pole observables in the space of the Warsaw basis operators spanned by $O_{Hf}$, $O_{WB}$, and $O_{T}$. Of course, the number of constraints is the same in the Higgs and Warsaw basis: in any basis,  the pole observables in \tref{EWPT_zpole} and \tref{EWPT_wpole} always constrain 21 linear combinations of Wilson coefficients.

\vspace{0.5cm}

Different flavor models lead to specific patterns of vertex corrections. In particular, they often impose relations between different $[\delta g]_{ij}$'s, reducing the number of free parameters. In the following we discuss some simple flavor structures for the effective operators, the resulting pattern of vertex corrections, and the constraints on the parameters of these scenarios.

\subsection{Flavor Universality}

The simplest flavor scenario is the one assuming an unbroken $U(3)_F^5$ flavor symmetry for the $D=6$ Lagrangian, as previously considered in Refs.~\cite{Pomarol:2013zra,Ellis:2014huj,Falkowski:2014tna}. This ansatz leads to flavor blind vertex corrections,
\beq
[\delta g^{Vf}_{L,R}]_{ij} = A^{Vf}_{L,R} \, \delta_{ij} \, .
\eeq
Among this eight dimensional parameter space only seven directions affect the pole observables at the linear level ($\delta g^{Wq}_R$ enters only quadratically, see above). 
In this case, instead of the leptonic Z-pole observables in  \tref{EWPT_zpole} and leptonic W branching fractions in \tref{EWPT_wpole},  
we use the corresponding observables determined under assumption of lepton flavor universality, see Table~1 of Ref.~\cite{Falkowski:2014tna}.
We find
\beq
\label{eq:flavorblind}
\hspace{-.7cm}
\left ( \begin{array}{c}
A^{W \ell}_L  \\
A^{Ze}_L  \\
A^{Ze}_R \\
A^{Zu}_L \\
A^{Zu}_R \\
A^{Zd}_L \\
A^{Zd}_R \\
\end{array} \right )  =
\left ( \begin{array}{c}
- 0.89 \pm    0.84 \\
- 0.20  \pm   0.23  \\
- 0.20   \pm 0.24    \\
-1.7 \pm  2.1   \\
 - 2.3    \pm   4.6  \\
2.8          \pm 1.5   \\
19.9         \pm  7.7
\end{array} \right )   \times 10^{-3}\, ,
\eeq
where the correlation matrix is given in \eref{rhoUni}  of \aref{rho21}.

\subsection{Alignment}

In the alignment scenario one assumes that the flavor structure of the different $\mathcal{O}_6$'s is aligned with the corresponding Yukawa matrix.  In more detail, the right-handed currents are aligned with $Y_{u,d}Y_{u,d}^\dagger$, while the left-handed ones with $Y_{u,d}^\dagger Y_{u,d}$.
For the latter, one has to specify whether these are aligned with the Yukawa matrix of the up sector (up-alignment) or the with the Yukawa matrix of the down sector (down-alignment).
In our basis, the vertex corrections then take the form
\beq
\delta g^{Z f}_R =  [\delta g^{Z f}_R]_{ii} \delta_{ij},
\eeq
for $f = u,d,e$, and
\beq
[\delta g^{Z u}_L]_{ij} = [\delta g^{Z u}_L]_{ii} \delta_{ij},  \quad  \delta g^{Z d}_L = \sum_k[\delta g^{Z d}_L]_{kk}V^*_{ki}V_{kj}
\qquad \text{(up-alignment)},
\eeq
or
\beq
[\delta g^{Z d}_L]_{ij} = [\delta g^{Z d}_L]_{ii} \delta_{ij}, \quad \delta g^{Z u}_L = \sum_k[\delta g^{Z u}_L]_{kk}V_{ik}V^*_{jk}\qquad \text{(down-alignment)}.
\eeq
Moreover, for the lepton sector,
\beq
[\delta g^{Z \nu}_L]_{ij} = [\delta g^{Z \nu}_L]_{ii} \delta_{ij},  \quad   \delta g^{Z e}_L = \sum_k [\delta g^{Z e}_L]_{kk} U^*_{ki} U_{kj}
\qquad \text{($\nu$-alignment)},
\eeq
or
\beq
[\delta g^{Z e}_L]_{ij} = [\delta g^{Z e}_L]_{ii} \delta_{ij}, \quad  \delta g^{Z \nu}_L = \sum_k [\delta g^{Z \nu}_L]_{kk} U_{ik} U_{jk}^* \qquad \text{($e$-alignment)},
\eeq
where the CKM matrix $V$ and the PMNS matrix  $U$ are taken from~\cite{Beringer:1900zz}. Clearly, the alignment hypothesis does not reduce the number of independent diagonal vertex corrections. The resulting constraints on the diagonal correction are found to be the same as in Eqs.~\eqref{eq:dgl}--\eqref{eq:dgZd}.

The off-diagonal couplings of the left-handed quarks are controlled by the non-universality in the diagonal vertex corrections. 
To leading order in the Wolfenstein parameter $\lambda_C$, one obtains for the up-alignment: 
\begin{align}
[\delta g^{Z d}_L]_{12} \simeq& \left( [\delta g^{Z d}_L]_{11} - [\delta g^{Z d}_L]_{22} \right) \lambda_C \, , \nnl
[ \delta g^{Z d}_L]_{13} \simeq& \left( [\delta g^{Z d}_L]_{33}-[\delta g^{Z d}_L]_{22}
+( [\delta g^{Z d}_L]_{11}-[\delta g^{Z d}_L]_{33} )(\rho-i \eta) \right) A \lambda_C^3 \, , \nnl
[\delta g^{Z d}_L]_{23} \simeq& \left( [\delta g^{Z d}_L]_{22} - [\delta g^{Z d}_L]_{33} \right) A \lambda_C^2 \, ,
\end{align}
and for the down-alignment: 
\begin{align}
[\delta g^{Z u}_L]_{12} \simeq& \left( [\delta g^{Z u}_L]_{22} - [\delta g^{Z u}_L]_{11} \right) \lambda_C \, , \nnl
[ \delta g^{Z u}_L]_{13} \simeq& \left( [\delta g^{Z u}_L]_{11}-[\delta g^{Z u}_L]_{22}
-( [\delta g^{Z u}_L]_{11}-[\delta g^{Z u}_L]_{33} )(\rho-i\eta) \right) A \lambda_C^3 \, , \nnl
[\delta g^{Z u}_L]_{23} \simeq& \left( [\delta g^{Z u}_L]_{33} - [\delta g^{Z u}_L]_{22} \right) A \lambda_C^2 \, .
\end{align}
with $A$, $\eta$ and $\rho$ are the other Wolfenstein parameters of the CKM matrix. Clearly, at the limit of universal diagonal vertex corrections the off-diagonal couplings vanish. Given the limits on the diagonal vertex corrections, we find that
\bea \label{eq:ZoffAlign}
\text{(up-alignment):}
&\!\!\!\! [\delta g^{Z d}_L]_{12} \lesssim 3 \times 10^{-2}\, ,  
\   [\delta g^{Z d}_L]_{13} \lesssim 7 \times 10^{-4}\, , 
\  [\delta g^{Z d}_L]_{23} \lesssim 2 \times 10^{-3}\, , 
\nnl
\text{(down-alignment):}
&\!\!\!\! [\delta g^{Z u}_L]_{12} \lesssim 1 \times 10^{-2}\, ,
\   [\delta g^{Z u}_L]_{13} \lesssim 5 \times 10^{-4}\, , 
\   [\delta g^{Z u}_L]_{23} \lesssim 3 \times 10^{-3}\, , 
\nnl
\text{($\nu$-alignment):}
&\!\!\!\! [\delta g^{Z e}_L]_{12} \lesssim 9 \times 10^{-4}\, , 
\  [\delta g^{Z e}_L]_{13} \lesssim 7 \times 10^{-4}\, , 
\  [\delta g^{Z e}_L]_{23} \lesssim 9 \times 10^{-4}\, , 
\eea
is allowed at 95\%~CL. We see that, in the down-alignment case, the allowed magnitude of $[\delta g^{Z u}_L]_{23}$ 
is just below the direct limit from $t \to Zc$ constraints, and may be probed by these searches in the forthcoming LHC run. Similarly, in the $\nu$-alignment case, the upper limits are not far from  the direct bounds on $Z$ lepton flavor violating decays,~\eref{ZLFV}.

Indirect constraints on the $Z$ off-diagonal couplings also arise from low-energy processes. These bounds are sensitive to the assumptions on the 4-fermion operators, and hence meaningful only in the absence of cancelation between the different FCNC contributions. Although not the scope of this paper, we analyze these constraints in~\aref{indirectZoff} in the alignment scenario for completeness. Currently, these are the only available observables which are sensitive to the off-diagonal couplings to $d,s,b,u$ and $c$. By comparing the bounds of~\eref{ZoffIndirectUp} and \eref{ZoffIndirectDw} to the allowed ranges of~\eref{ZoffAlign} we conclude the following: the indirect bound on $[\delta g^{Z u}_L]_{12}$ from charm mixing measurements is about a factor of 40 stronger than the allowed range in down alignment scenario. In case of up-alignment, the bound from Kaon-mixing on $[\delta g^{Z d}_L]_{12}$ is stronger by two orders of magnitude, while the bounds from $B_{d,(s)}\to \mu^+\mu^-$ on $[\delta g^{Z d}_L]_{13(23)}$ are stronger only by an order of magnitude (factor of few) from the allowed range.

\subsection{Minimal Flavor Violation}

Another extreme solution to the NP flavor puzzle is the one of MFV. The assumption of MFV states that the SM Lagrangian, as well as any NP interactions, formally respect a global $SU(3)^5_F$ flavor symmetry. Under this ansatz, the SM fermions transforms in the fundamental representation of the corresponding $SU(3)_F$, while the Yukawas are spurion fields following a bi-fundamental transformation low. In the MFV scenario the global $SU(3)^5_F$ symmetry is then broken by the expectation values of the Yukawa spurions, given by the fermion masses and mixing parameters. The effective interactions induced by the heavy states should then be formally invariant under this symmetry. In this section we discard neutrino masses and hence have no effect arising from the leptonic mixing parameters.

Imposing the MFV ansatz on the effective Lagrangian, to leading order in the spurions, the vertex correction receive the following contributions: 
\bea
\delta g_L^{Zu} &=& A_L^{Zu} \im + \tilde{B}_L^{Zu} Y_u^\dagger Y_u + \tilde{C}_L^{Zu} Y_d^\dagger Y_d  \nnl
&=&A_L^{Zu}\delta_{ij}
+\frac{1}{2}B_L^{Zu}\left( \frac{m_{u_i}^2}{m_t^2}\delta_{ij}+\sum_{d_k}\frac{m_{d_k}^2}{\tilde{m}_b^2}V_{ik}V^*_{jk} \right)
+\frac{1}{2}C_L^{Zu}\left( \frac{m_{u_i}^2}{m_t^2}\delta_{ij}-\sum_{d_k}\frac{m_{d_k}^2}{\tilde{m}_b^2}V_{ik}V^*_{jk} \right),
\nnl
\delta g_L^{Zd} &= &   A_L^{Zd} \im + \tilde{B}_L^{Zd} Y_d^\dagger Y_d + \tilde{C}_L^{Zd} Y_u^\dagger Y_u \nnl
&=&A_L^{Zd}\delta_{ij}
+\frac{1}{2}B_L^{Zd}\left( \frac{m_{d_i}^2}{m_b^2}\delta_{ij}+\sum_{u_k}\frac{m_{u_k}^2}{\tilde{m}_t^2}V^*_{ki}V_{kj} \right)
+\frac{1}{2}C_L^{Zd}\left( \frac{m_{d_i}^2}{m_b^2}\delta_{ij}-\sum_{u_k}\frac{m_{u_k}^2}{\tilde{m}_t^2}V^*_{ki}V_{kj}\right),
\nnl
\delta g_R^{Zu} &= &   A_R^{Zu} \im + \tilde B_R^{Zu} Y_u Y_u^\dagger=\left(A_R^{Zu}+B_R^{Zu}\frac{m_{u_i}^2}{m_t^2}\right)\delta_{ij},
\nnl
\delta g_R^{Zd} &= &  A_R^{Zd} \im + \tilde B_R^{Zd} Y_d Y_d^\dagger=\left(A_R^{Zd}+B_R^{Zd}\frac{m_{d_i}^2}{m_b^2}\right)\delta_{ij},
\nnl
\delta g_L^{W\ell} &= & A_L^{W \ell} \im + \tilde B_L^{W \ell} Y_e^\dagger Y_e=\left(A_L^{W\ell}+B_L^{W \ell}\frac{m_{e_i}^2}{m_\tau^2}\right)\delta_{ij},
\nnl
\delta g_L^{Ze} &= &  A_L^{Ze} \im +  \tilde B_L^{Ze} Y_e^\dagger Y_e=\left(A_L^{Ze}+B_L^{Ze}\frac{m_{e_i}^2}{m_\tau^2}\right)\delta_{ij},
\nnl
\delta g_R^{Ze} &= &  A_R^{Ze} \im +  \tilde B_R^{Ze} Y_e Y_e^\dagger=\left(A_R^{Ze}+B_R^{Ze}\frac{m_{e_i}^2}{m_\tau^2}\right)\delta_{ij},
\eea
where we take the fermion masses at $m_Z$ from~\cite{Xing:2011aa}, and use $\tilde{m}_b^2\equiv\sum_k{V_{3k}V^*_{3k}}m_{d_k}^2\simeq m_b^2$, $\tilde{m}_t^2\equiv\sum_k{V_{k3}V^*_{k3}}m_{u_k}^2\simeq m_t^2$.  $B^{Zu}_R$ is very weakly constrained because  $[\delta g^{Zu}_R]_{33}$ is not bounded. In addition, both $C^{Zu}_L$ and $C^{Zd}_L$ do not modify the couplings to the third generation and hence they are very weakly constrained by the data. We neglect these in our numerical fit. The vertex corrections are now parameterized by 14 parameters, with contributions which are correlated across different observables.
We find that, under the MFV assumption, the limits on the expansion coefficients are given by 
\beq
\label{eq:mfv}
\hspace{-.7cm}
\left ( \begin{array}{c}
A^{W \ell}_L  \\
B^{W \ell}_L  \\
A^{Ze}_L  \\
B^{Ze}_L  \\
A^{Ze}_R \\
B^{Ze}_R \\
A^{Zu}_L \\
B^{Zu}_L \\
A^{Zd}_L \\
B^{Zd}_L \\
A^{Zu}_R \\
A^{Zd}_R \\
B^{Zd}_R \\
\end{array} \right )  =
\left ( \begin{array}{c}
- 1.2 \pm    0.4 \\ 
- 3.2 \pm    1.2 \\  
- 0.021  \pm   0.024  \\ 
0.039   \pm 0.062    \\  
- 0.031  \pm   0.025  \\   
0.073   \pm 0.066    \\ 
-0.19  \pm 0.31   \\  
 - 0.1     \pm   3.8  \\ 
 0.20  \pm  0.54   \\ 
 0.12     \pm   0.57  \\ 
 -0.26  \pm  0.50   \\ 
1.6         \pm 2.7   \\ 
0.7        \pm  2.9 
\end{array} \right )   \times 10^{-2}\, .
\eeq
The correlation matrix is given in~\eref{rhoMFV} of~\aref{rho21}. The off-diagonal terms are extremely suppressed by the fermion masses and the CKM elements. In particular, the upper possible value for $[\delta g^{Zu}_L]_{23}\lesssim 1.6\times 10^{-3}$ is an order of magnitude below its current experimental bound, see~\eref{tzc}.

Higher order corrections in the MFV expansion might modify the relations between the couplings to different generation in each sector~\cite{Kagan:2009bn}. Yet again, due to the $m^2$ suppression the significant bounds arise only from the coupling to the third generation. Hence, including these higher contributions is equivalent to a redefinition of the various $B$'s, and can be made straightforwardly.

\subsection{Anarchic vector-like fermions}

Another common flavor ansatz is the idea of mixing between the SM fermions and heavy vector-like states with an anarchic flavor structure. In the anarchic scenario one assumes the absence of any direct couplings between the SM fields and the Higgs doublet. Instead, the masses and mixing are generated solely via this mixing, which induce effectively the familiar Yukawa interactions. A similar phenomenology is retained in the anarchic partial compositeness scenario, which can be realized in composite Higgs models or in the warped extra dimension~\cite{ArkaniHamed:1999dc, Gherghetta:2000qt, Huber:2000ie,Agashe:2004cp}. One can further assume that the hierarchic flavor structure is encoded entirely in the mixing parameters, rather than in the vector-like sector itself. Under this assumption, the effective Yukawa matrices are determined by
\beq
[Y_f]_{ij} = \lambda^{f_R}_i [\tilde Y_f]_{ij} \lambda^{f_L}_j,
\eeq
where $\lambda^{f}$ is the mixing strength between the vector-like fermions and the SM fields, assumed to obey $\lambda^{d_L}=\lambda^{u_L}$, and the anarchic ansatz states that $\tilde Y_f$ are random matrices of order one. The mixing parameters are determined, up to order one factors, by the observed masses and mixing angles~\cite{Huber:2000ie},
\bea
&&\frac{m_{u_i}}{v} \sim \lambda_i^u\lambda_i^q \, , \;\;\;
\frac{m_{d_i}}{v} \sim \lambda_i^d\lambda_i^q \, , \;\;\;
V_{ij} \sim \frac{\lambda_i^q}{\lambda_j^q}\, , \;\;\;{\rm ~for~}{i<j}\,.
\eea
As a convenient choice we take $\lambda_3^q=\mathcal{O}(1)$ which in turn dictate the order of all other parameters in the quark sector. The same parameters also set the order of magnitude of the vertex corrections, obeying
\begin{align}\label{eq:comp}
[\delta g_L^{Zu}]_{ij}
&=   A_1\delta_{ij}+\left[\tilde{B}^{Zu}_L\right]_{ij}  \lambda_i^q \lambda_j^q
=   A_1\delta_{ij}+\left[B^{Zu}_L\right]_{ij}  V_{ib}V_{jb} \, , \nnl
[\delta g_L^{Zd}]_{ij}
&=   A_2\delta_{ij}+\left[\tilde{B}^{Zd}_L\right]_{ij}  \lambda_i^q \lambda_j^q
= A_2\delta_{ij}+\left[B^{Zd}_L\right]_{ij}  V_{ib}V_{jb} \, , \nnl
[\delta g_R^{Zu}]_{ij}
&=  2(A_1+A_2)\delta_{ij}+\left[\tilde{B}^{Zu}_R\right]_{ij} \lambda_i^u \lambda_j^u
=  2(A_1+A_2)\delta_{ij}+\left[B^{Zu}_R\right]_{ij} \frac{m_{u_i} m_{u_j}}{v^2}\frac{1}{V_{ib}V_{jb}}  \, , \nnl
[\delta g_R^{Zd}]_{ij}
&= -(A_1+A_2)\delta_{ij}+\left[\tilde{B}^{Zd}_R\right]_{ij} \lambda_i^d \lambda_j^d
= -(A_1+A_2)\delta_{ij}+\left[B^{Zd}_R\right]_{ij} \frac{m_{d_i} m_{d_j}}{v^2}\frac{1}{V_{ib}V_{jb}} \, , \nnl
[\delta g_L^{W\ell}]_{ij} &= (A_1-A_2)\delta_{ij} \, , \nnl
[\delta g_L^{Ze}]_{ij} &= -(2A_1+A_2)\delta_{ij}  \, , \nnl
[\delta g_R^{Ze}]_{ij} &= -3(A_1+A_2)\delta_{ij}  \,.
\end{align}
One could add the corresponding leptonic flavor dependent contributions in a similar form as the ones in the quark sector. However, assuming a common NP scale for all sectors, the corresponding $\lambda^\ell$ and $\lambda^e$ are expected to be suppressed by the small lepton masses. For instance, taking
\begin{align}
\frac{m_{e_i}}{v} \sim \lambda_i^e\lambda_i^\ell \, , \;\;\;\lambda_i^\ell \sim \lambda_3^e \sim \sqrt{m_\tau/v}
\end{align}
will generate the required fermionic mass hierarchy and leptonic mixing structure. The overall effect of these parameters on the vertex corrections is negligible due to the overall mass suppression they exhibit.

As a meaningful result we quote in the following only the bounds for which the different generation are not split by more than two orders of magnitude and set all the other couplings to zero. The resulting bounds are
\beq
  \label{eq:composite}
\left ( \begin{array}{c}
A_1  \\
A_2
\end{array} \right )  =
\left ( \begin{array}{c}
0.3 \pm 2.0 \\
-0.4\pm2.2
\end{array} \right )   \times 10^{-4}\, ,
\hspace{.7cm}
\left ( \begin{array}{c}
\left[B^{Zu}_L\right]_{22}  \\
\left[B^{Zu}_L\right]_{33}  \\
\left[B^{Zd}_L\right]_{33}  \\
\left[B^{Zu}_R\right]_{22} 
\end{array} \right )  =
\left ( \begin{array}{c}
-39\pm130 \\
-0.7\pm3.8    \\
-0.043\pm0.067    \\
-40\pm120    
\end{array} \right )   \times 10^{-2}\, ,\eeq
with the corresponding correlation matrix given in~\eref{rhoPC} of~\aref{rho21}. This class of models leads to an interesting flavor phenomenology~\cite{Agashe:2004cp,Agashe:2006wa,Agashe:2004ay,Davidson:2007si}. However, the large contribution to $Z\to \bar{b}_L b_L$ pushes the NP scale to the scale of order $4\,$TeV~\cite{Agashe:2005dk}. In Ref.~\cite{Agashe:2006at}, it was shown that a custodial symmetry can protect the $Z\to \bar{b}_L b_L$  vertex, resulting in a valid lower NP scale. Note that the parametric suppression of the right-handed currents with custodial symmetry is slightly different, $\sim (m_i m_j /v^2)V_{ib}/V_{jb}$.  A detailed discussion on rare $K$ and $B$ decays in custodial protected models can be found in~\cite{Blanke:2008yr}, while a discussion on top flavor violating decays can be found in~\cite{Azatov:2014lha}.

Three comments are in order. 
First, we note the $A_{1,2}$ universal parts in~\eref{comp} which arise from the oblique contributions to the vertex corrections in our basis.
These encode the effect of the usual $S$ and $T$ oblique parameters, that typically arise in scenarios of this kind. Second, we stress that in our analysis we assume no accidental cancelation between different contributions to the vertex corrections. Furthermore, in concrete models of partial compositeness stronger limits may arise from other effects than the vertex corrections, \textit{ e.g.} from 4-quark operators induced by heavy gluon exchange~\cite{Csaki:2008zd}.

\section{Conclusions}
\setcounter{equation}{0}
\label{sec:con}

In this paper we derived model-independent constraints on the $D=6$ Lagrangian from the $Z$ and $W$ pole observables.
These observables constrain the corrections to the $W$ boson mass and to the $W$ and $Z$ boson interactions with SM fermions.
Our main result is displayed in Eqs.~\eqref{eq:dgl}--\eqref{eq:dgZd}, from which the following conclusions can be drawn.
\bi
\item
Flavor diagonal leptonic couplings are robustly constrained.
The limits are most stringent on the $Z$ couplings to charged leptons, where the deviations from the SM are at most $\cO(10^{-4})-\cO(10^{-3})$. 
Leptonic couplings of $W$ (and by gauge symmetry of the effective Lagrangian, also $Z$ couplings to neutrinos) are somewhat less constrained, at the level of $\cO(10^{-2})$.
Moreover, one can also constrain flavor off-diagonal $Z$ couplings to charged leptons at the level of $\cO(10^{-3})-\cO(10^{-2})$.
\item
For quark couplings,  the limits depend a lot on the flavor.
Couplings to the bottom and charm quarks are still fairly well constrained, at the level of $\cO(10^{-2})$.
Constraints on other quark couplings are weaker, and $\cO(1)$ deviations are allowed in some cases.
Constraints on off-diagonal $Z$ couplings involving the top quark are currently  $\cO(10^{-1})$. We emphasize that for case of large Wilson coefficients (translated to large vertex corrections) the validity of the EFT expansion should be verified and that the new physics scale itself should be well above the EW one. 
\ei
The above bound on the vertex corrections can be translated to the bound on the scale $\Lambda$ suppressing the respective dimension six operator:  $\Lambda \gtrsim 5  \sqrt{10^{-3}/\delta g }$~TeV. 

Our results have important consequences for ongoing searches for physics beyond the SM.
In principle, the vertex corrections could affect the total rate and differential distributions of numerous processes at the LHC.
The limits we provide imply model-independent bounds on the magnitude of such effects.
For example, for Higgs boson decays  to four leptons via intermediate gauge bosons, the effect of vertex corrections will be difficult to observe,  and can be safely neglected in current LHC Higgs  analyses.

At the same time, we have shown that certain electroweak couplings are poorly or not-at-all constrained in a model independent way.
One blatant example is the $Z$ boson coupling to right-handed top quarks. Currently, the observables sensitive to this coupling (such as $b \to s \gamma$, or $ttZ$ associated production) depend also on other dimension-6 operators (4-fermion couplings, dipole couplings of the top quark), which makes difficult extracting model-independent constraints. A dedicate analysis for the EW and rare $K$ and $B$ decays on $ttZ$ vertex coupling can be found in~\cite{Brod:2014hsa}, while direct and indirect constraints on top dipole moments are given  in~\cite{Kamenik:2011dk}.
Precision measurements of  the $Z t_R t_R$ coupling is  one of the strongest  motivations for building a high-energy $e^+e^-$ collider~\cite{Amjad:2013tlv,Richard:2014upa}.

Next, the pole observables alone provide no constraints on the flavor off-diagonal $Z$ couplings to light quarks.
While these couplings affect meson mixing, their contribution is entangled with that of four-quark operators.
Therefore a more general analysis that includes these operators is in order to establish model-independent bounds on off-diagonal quark couplings.
Non-trivial limits from the pole observables can be obtained in the context of particular flavor models, where the off-diagonal couplings are correlated with the diagonal ones.

Finally, the $Z$ boson couplings to light quarks are presently only weakly constrained. 
These couplings are probed by multiple high-precision measurements, for example, by atomic parity violation, parity-violating electron scattering, fermion pair production in LEP-2, and meson decays. However, these processes involve an off-shell  $Z$ boson exchange, and as a consequence they are also sensitive to four-fermion operators involving electrons and quarks. Again, a more general analysis that includes these operators is needed in order to establish model-independent constraints using these processes.
The $Z$ boson couplings to light quarks can also be probed in hadron colliders.
Indeed, it was demonstrated that hadron colliders can achieve a decent precision to measure electroweak parameters, in particular $\sin^2 \theta_W$~\cite{Chatrchyan:2011ya,Abazov:2014jti}.  Model independent measurements of $Z$ boson couplings to up and down quarks, as done in Ref.~\cite{Abazov:2011ws}, can be repeated at the LHC and with the full Tevatron dataset.

\section*{Acknowledgements}

We thank Wolfgang Altmannshofer, David Marzocca, Yossi Nir, Gilad Perez, Francesco Riva,  and Emmanuel Stamou for useful discussions.
We also thank Luigi Rolandi for suggesting to include Drell-Yan measurements in hadron collider, and Fabio Maltoni for a comment on the model dependence of constraints from the $ttZ$ production at the LHC.
AF~is supported by the ERC Advanced Grant Higgs@LHC.

\appendix
\renewcommand{\theequation}{\Alph{section}.\arabic{equation}}

\section{Results in other bases}
\label{app:otherbases}

In this appendix we discuss the relation between the vertex and mass corrections in our effective Lagrangian, and the Wilson coefficients of $SU(3)\times SU(2)\times U(1)$ invariant  $D=6$ operators in two different bases used in the literature.

\subsection{Warsaw basis}
\label{app:warsaw}
\setcounter{equation}{0}

\begin{table}[hb]
\begin{center}
\small
\begin{minipage}[t]{4cm}
\renewcommand{\arraystretch}{1.5}
\begin{tabular}[t]{c|c}
\multicolumn{2}{c}{$H^4 D^2$ and $H^6$} \\
\hline
$O_{H}$ & $ \left [ \partial_\mu (H^\dag H) \right ]^2$ \\
$O_{T}$   & $ \left (H^\dagger {\overleftrightarrow { D_\mu}} H \right)^2 $  \\
$O_{6H}$       & $(H^\dag H)^3$
\end{tabular}
\end{minipage}
\begin{minipage}[t]{5cm}
\renewcommand{\arraystretch}{1.5}
\begin{tabular}[t]{c|c}
\multicolumn{2}{c}{$f^2 H^3$} \\
\hline
$O_{e}$           & $- (H^\dag H - {v^2 \over 2}) \bar e H^\dagger \ell$ \\
$O_{u}$          & $- (H^\dag H- {v^2 \over 2}) \bar u  \widetilde H^\dagger  q$ \\
$O_{d}$           & $- (H^\dag H- {v^2 \over 2}) \bar d H^\dagger q$\\
\end{tabular}
\end{minipage}
\begin{minipage}[t]{5cm}
\renewcommand{\arraystretch}{1.5}
\begin{tabular}[t]{c|c}
\multicolumn{2}{c}{$V^3 D^3$} \\
\hline
$O_{3G}$                & $g_s^3 f^{abc} G_{\mu \nu}^{a} G_{\nu\rho}^b G_{\rho\mu}^c $ \\
$O_{\widetilde {3G} }$          & $g_s^3 f^{abc} \widetilde  G_{\mu \nu}^{a} G_{\nu\rho}^b G_{\rho\mu}^c $ \\
$O_{3W}$                & $g^3 \epsilon^{ijk} W_{\mu \nu}^i W_{\nu \rho}^j W_{\rho\mu}^k$ \\
$O_{\widetilde {3W} }$          & $g^3 \epsilon^{ijk} \widetilde  W_{\mu \nu}^i W_{\nu \rho}^j W_{\rho\mu}^k$ \\
\end{tabular}
\end{minipage}
\vspace{0.25cm}
\begin{minipage}[t]{5cm}
\renewcommand{\arraystretch}{1.5}
\begin{tabular}[t]{c|c}
\multicolumn{2}{c}{$V^2H^2$} \\
\hline
$O_{GG}$     & $ g_s^2 H^\dag H\, G^a_{\mu\nu} G^a_{\mu\nu}$ \\
$O_{\widetilde {GG} }$         & $ g_s^2  H^\dag H\, \widetilde G^a_{\mu\nu} G^a_{\mu\nu}$ \\
$O_{WW}$     & $g_L^2 H^\dag H\, W^i_{\mu\nu} W^i_{\mu\nu}$ \\
$O_{\widetilde {WW}}$         & $g_L^2 H^\dag H\, \widetilde W^i_{\mu\nu} W^i_{\mu\nu}$ \\
$O_{BB}$     & $ g_Y^2H^\dag H\, B_{\mu\nu} B_{\mu\nu}$ \\
$O_{\widetilde {BB}}$         & $g_Y^2  H^\dag H\, \widetilde B_{\mu\nu} B_{\mu\nu}$ \\
$O_{WB}$     & $ g_L g_Y H^\dag \sigma^i H \, W^i_{\mu\nu} B_{\mu\nu}$ \\
$O_{\widetilde {W B}}$         & $g_L g_Y H^\dag \sigma^i H \, \widetilde W^i_{\mu\nu} B_{\mu\nu}$
\end{tabular}
\end{minipage}
\begin{minipage}[t]{5cm}
\renewcommand{\arraystretch}{1.5}
\begin{tabular}[t]{c|c}
\multicolumn{2}{c}{$f^2 H^2 D$} \\
\hline
$O_{H \ell }$      & $i  \bar \ell \bar \sigma_\mu  \ell H^\dagger \overleftrightarrow {D_\mu} H$\\
$O_{H \ell}'$      & $i  \bar \ell \sigma^i \bar \sigma_\mu  \ell  H^\dagger \sigma^i \overleftrightarrow {D_\mu} H $\\
$O_{H e}$            & $i e^c \sigma_\mu \bar   e^c  H^\dagger  \overleftrightarrow {D_\mu} H$\\
$O_{H q}$      & $ i  \bar q  \bar \sigma_\mu   q H^\dagger  \overleftrightarrow {D_\mu} H  $\\
$O_{H q}'$      & $i \bar q \sigma^i  \bar \sigma_\mu  q H^\dagger \sigma^i \overleftrightarrow {D_\mu} H$\\
$O_{H u}$            & $ i u^c \sigma_\mu   \bar u^c  H^\dagger  \overleftrightarrow {D_\mu} H $\\
$O_{H d}$            & $i  d^c \sigma_\mu \bar d^c  H^\dagger  \overleftrightarrow {D_\mu} H $\\
$O_{H u d}$  & $i   u^c  \sigma_\mu  \bar d^c  \tilde H^\dagger {D_\mu} H$ \\
\end{tabular}
\end{minipage}
\begin{minipage}[t]{5.2cm}
\renewcommand{\arraystretch}{1.5}
\begin{tabular}[t]{c|c}
\multicolumn{2}{c}{$f^2 V H D$} \\
\hline
$O_{eW}$      & $g_L \bar \ell \sigma_{\mu\nu} \bar e^c \sigma^i H W_{\mu\nu}^i$ \\
$O_{eB}$        & $g_Y \bar \ell \sigma_{\mu\nu}  \bar  e^c H B_{\mu\nu}$ \\
$O_{uG}$        & $g_s \bar q \sigma_{\mu\nu} T^a  \bar  u^c \widetilde H \, G_{\mu\nu}^a$ \\
$O_{uW}$        & $g_L \bar q  \sigma_{\mu\nu}  \bar  u^c \sigma^i \widetilde H \, W_{\mu\nu}^i$ \\
$O_{uB}$        & $g_Y \bar q  \sigma_{\mu\nu}  \bar  u^c  \widetilde H \, B_{\mu\nu}$ \\
$O_{dG}$        & $g_s \bar q \sigma_{\mu\nu} T^a  \bar  d^c H\, G_{\mu\nu}^a$ \\
$O_{dW}$         & $g_L \bar q \sigma_{\mu\nu} \bar   d^c \sigma^i H\, W_{\mu\nu}^i$ \\
$O_{dB}$        & $g_Y \bar q \sigma_{\mu\nu}  \bar  d^c  H\, B_{\mu\nu}$
\end{tabular}
\end{minipage}
\vspace{0.25cm}

\begin{minipage}[t]{5.5cm}
\renewcommand{\arraystretch}{1.5}
\begin{tabular}[t]{c|c}
\multicolumn{2}{c}{$(\bar LL)(\bar LL)$ and $(\bar LR)(\bar L R)$} \\
\hline
$O_{\ell\ell}$        & $(\bar \ell \bar \sigma_\mu \ell)(\bar \ell \bar \sigma_\mu \ell)$ \\
$O_{qq}$  & $(\bar q \bar \sigma_\mu q)(\bar q \bar \sigma_\mu q)$ \\
$O_{qq}'$  & $(\bar q \bar \sigma_\mu \sigma^i q)(\bar q \bar \sigma_\mu \sigma^i q)$ \\
$O_{\ell q}$                & $(\bar \ell \bar \sigma_\mu \ell)(\bar q \bar \sigma_\mu q)$ \\
$O_{\ell q}'$                & $(\bar \ell \bar \sigma_\mu \sigma^i \ell)(\bar q \bar \sigma_\mu \sigma^i q)$  \\
$O_{quqd}$ & $(u^c q^j) \epsilon_{jk} (d^c q^k)$ \\
$O_{quqd}'$ & $(u^c T^a q^j) \epsilon_{jk} (d^c T^a q^k)$ \\
$O_{\ell equ}$ & $(e^c  \ell^j) \epsilon_{jk} (u^c q^k)$ \\
$O_{\ell equ}'$ & $(e^c \bar \sigma_{\mu\nu} \ell^j ) \epsilon_{jk} ( u^c \bar \sigma^{\mu\nu} q^k  )$ \\
$O_{\ell edq}$ & $(\bar \ell  \bar e^c)(d^c q)$
\end{tabular}
\end{minipage}
\begin{minipage}[t]{5.25cm}
\renewcommand{\arraystretch}{1.5}
\begin{tabular}[t]{c|c}
\multicolumn{2}{c}{$(\bar RR)(\bar RR)$} \\
\hline
$O_{ee}$               & $( e^c \sigma_\mu \bar e^c)( e^c \sigma_\mu \bar e^c)$ \\
$O_{uu}$        & $(u^c \sigma_\mu \bar u^c)(u^c \sigma_\mu \bar u^c)$ \\
$O_{dd}$        & $(d^c \sigma_\mu \bar d^c)(d^c \sigma_\mu \bar d^c)$ \\
$O_{eu}$                      & $(e^c \sigma_\mu \bar e^c)(u^c \sigma_\mu \bar u^c)$ \\
$O_{ed}$                      & $(e^c \sigma_\mu \bar e^c)(d^c \sigma_\mu \bar d^c)$ \\
$O_{ud}$                & $(u^c \sigma_\mu \bar u^c)(d^c \sigma_\mu \bar d^c)$ \\
$O_{ud}'$                & $(u^c \sigma_\mu T^a \bar u^c)(d^c \sigma_\mu T^a \bar d^c)$ \\
\end{tabular}
\end{minipage}
\begin{minipage}[t]{4.75cm}
\renewcommand{\arraystretch}{1.5}
\begin{tabular}[t]{c|c}
\multicolumn{2}{c}{$(\bar LL)(\bar RR)$} \\
\hline
$O_{\ell e}$               & $(\bar  \ell \bar \sigma_\mu \ell)(e^c \sigma_\mu \bar  e^c )$ \\
$O_{\ell u}$               & $(\bar  \ell \bar \sigma_\mu \ell)(u^c  \sigma_\mu \bar  u^c )$ \\
$O_{\ell d}$               & $(\bar \ell \bar \sigma_\mu \ell)(d^c  \sigma_\mu \bar  d^c )$ \\
$O_{qe}$               & $(\bar q\bar  \sigma_\mu q)(e^c  \sigma_\mu \bar  e^c )$ \\
$O_{qu}$         & $(\bar  q\bar  \sigma_\mu q)(u^c  \sigma_\mu \bar  u^c )$ \\
$O_{qu}'$         & $(\bar q \bar \sigma_\mu T^a q)(u^c  \sigma_\mu T^a \bar  u^c )$ \\
$O_{qd}$ & $(\bar q \bar \sigma_\mu q)(d^c  \sigma_\mu \bar  d^c )$ \\
$O_{qd}'$ & $(\bar q \bar \sigma_\mu T^a q)(d^c  \sigma_\mu T^a \bar  d^c )$\\
\end{tabular}
\end{minipage}
\end{center}
\caption{\label{tab:warsaw}
Dimension six operators in the Warsaw basis~\cite{Grzadkowski:2010es}.
}
\end{table}

We consider the effective Lagrangian  $\cL_{\rm eff}^{\rm WB} = {\cal L}^ {\rm SM}  +  {1 \over v^2}  \sum_i c_i \mathcal{O}_{6,i}^{\rm WB}$,
where a complete non-redundant basis  of $D=6$ operators $\mathcal{O}_{6,i}^{\rm WB}$ is given in~\tref{warsaw}.
This basis is, up to small modifications, the same as in Ref.~\cite{Grzadkowski:2010es,Alonso:2013hga}, often referred to as the {\em Warsaw basis}.\footnote{%
The normalization of operators and notation are different than in the original references.
We replaced the operator $|H^\dagger D_\mu H|^2$ by $(H^\dagger D_\mu H - D_\mu H^\dagger H)^2$.
For Yukawa-type operators $O_f$ we subtracted $v^2$ so that these operators do not contribute to off-diagonal mass terms.
This way we avoid tedious rotations of the fermion fields to bring them back to the mass eigenstate basis.
Starting with the Yukawa couplings $- H \bar f'_R (Y_f'  + c_f' H^\dagger H/v^2) f_L'$ we can bring them to the form in ~\tref{warsaw} by defining
$f_{L,R}' =U_{L,R} f_{L,R}$, $c_f = U_R^\dagger c_f' U_L$, $Y_f  = U_R^\dagger (Y_f' + c_f'/2) U_L$, where $U_{L,R}$ are unitary rotations to the mass eigenstate basis.
}
In order to relate the two descriptions, we need to bring $\cL_{\rm eff}^{\rm WB}$ to the same form  as the effective Lagrangian considered in \sref{effl}.
In particular, we need to get rid of the kinetic mixing and non-canonical normalization induced by $\mathcal{O}_{6,i}^{\rm WB}$.
This is achieved by application of equations of motion, and field and coupling redefinitions, as described in Ref.~\cite{HXSWGbasis}.
When the dust settles, the shift of the $W$ boson mass is given by
\beq
\label{eq:D6_dm}
\delta m  =   {1 \over g_L^2 - g_Y^2} \left [ -  g_L^2  g_Y^2 c_{WB}   + g_L^2 c_T  -  g_Y^2 \delta v \right ],
\eeq
where $\delta v = ([c'_{H \ell}]_{11} +  [c'_{H \ell}]_{22})/2   + [c_{\ell \ell}]_{1221}/4$. 
The  leptonic vertex corrections are given by 
\bea
\delta g^{W \ell}_L & = &   c'_{H \ell} + f(1/2,0) - f(-1/2,-1),
\nnl
\delta g^{Z \nu }_L & = &    {1 \over 2} \left(c'_{H\ell} - c_{H\ell}\right)   +  f(1/2,0),
\nnl
\delta g^{Ze}_L & = &    - {1 \over 2} \left(c'_{H\ell}+c_{H\ell}\right) + f(-1/2, -1),
\nnl
\delta g^{Ze}_R & = &    - {1\over 2} c_{He}   +  f(0, -1),
\eea
where
\beq
f(T^3,Q) =  \im \left [- Q  c_{WB} {g_L^2 g_Y^2 \over g_L^2 - g_Y^2} + \left (c_T  - \delta v \right ) \left ( T^3 + Q {g_Y^2 \over g_L^2 - g_Y^2} \right ) \right ]. 
\eeq
Finally, the shifts of the SM $W$ and $Z$ boson couplings to quarks are given by
\bea
\delta g^{Wq}_L & = &  c'_{H q} V + f(1/2,2/3)  V  - f(-1/2,-1/3)  V ,
\nnl
\delta g^{W q}_R & = &   c_{Hud} ,
\nnl
\delta g^{Zu}_L & = &   {1 \over 2}  \left(c'_{Hq} - c_{Hq}\right) + f(1/2,2/3),
\nnl
\delta g^{Zd}_L & = &    -\frac{1}{2}V^\dagger \left(c'_{Hq}+c_{Hq}\right)V   + f(-1/2,-1/3),
\nnl
\delta g^{Zu}_R & = &    - {1\over 2} c_{Hu}   +  f(0,2/3),
\nnl
\delta g^{Zd}_R & = &    - {1\over 2} c_{Hd}  +  f(0,-1/3).
\eea
We can insert these relation into the global $\chi^2$ functions, so as to obtain constraints on the Wilson coefficients in the Warsaw basis.
Clearly, the vertex corrections constrained by pole observables map to a combination of a {\em larger} number of the Wilson coefficients $c_i$.
Therefore, only certain combinations of the latter can be constrained by the pole observables.
We define
\bea
\label{eq:EWPT_chats}
\left[\hat c_{H\ell}' \right]_{ij}  &=& \left[c_{H\ell}' \right]_{ij}  + \left ( g_L^2 c_{WB}  - {g_L^2 \over  g_Y^2} c_T  \right )\delta_{ij},
\nnl
\left[\hat c_{H\ell}\right]_{ij}   &=& \left[c_{H\ell} \right]_{ij}  -   c_T  \delta_{ij} ,
\nnl
\left[\hat  c_{He} \right]_{ij} & = & \left[ c_{He} \right]_{ij} -  2 c_T \delta_{ij} ,
\nnl
\left[\hat c_{Hq}' \right]_{ij} &=& \left[c_{Hq}'  \right]_{ij} + \left ( g_L^2 c_{WB}    - {g_L^2 \over g_Y^2 } c_T \right ) \delta_{ij},
\nnl
\left[\hat c_{Hq} \right]_{ij} &=& \left[c_{Hq}  \right]_{ij} + {1 \over 3} c_T   \delta_{ij},
\nnl
\left[\hat c_{Hu} \right]_{ij} &=&  \left[c_{Hu}  \right]_{ij} + {4 \over 3} c_T  \delta_{ij}  ,
\nnl
\left[\hat c_{Hd} \right]_{ij} &=&  \left[c_{Hd}  \right]_{ij} - { 2 \over 3 }c_T  \delta_{ij}. 
\eea
The pole observable  constrain all diagonal elements of $\hat c$ except for $[\hat{c}_{Hu}]_{33}$.

For these combinations, we obtain the following  central values and 1-sigma errors:
\beq
[c_{\ell\ell}]_{1221}  = (4.8 \pm 1.6) \times 10^{-2},
\nonumber 
\eeq
\beq
\left[\hat c'_{H\ell}\right]_{ii} = \bvec  -1.09  \pm 0.64 \\ -1.45 \pm 0.59 \\ 1.86 \pm 0.79 \evec  \times 10^{-2},
\quad
\left[\hat c_{H\ell}\right]_{ii}  = \bvec   1.03 \pm 0.63 \\  1.31 \pm 0.62  \\  -2.01 \pm 0.80 \evec  \times 10^{-2},
\nonumber
\eeq
\beq
\left[\hat c_{He}\right]_{ii}  = \bvec   0.22 \pm 0.66 \\  -0.6 \pm 2.6 \\  -1.3 \pm 1.3 \evec  \times 10^{-3},
\nonumber
\eeq
\beq
\left[\hat  c'_{Hq}\right]_{ii}  = \bvec  0.1 \pm 2.7 \\ -1.2 \pm 2.8 \\ -0.7 \pm 3.8 \evec  \times 10^{-2},
\quad
\left[\hat  c_{Hq}\right]_{ii}  = \bvec  1.8 \pm 7.1  \\ -0.8 \pm 2.9 \\  0.0 \pm 3.8 \evec \times 10^{-2},
\nonumber
\eeq
\beq
\label{eq:results_warsaw}
\left[\hat  c_{Hu}\right]_{ii}  = \bvec  -3 \pm 10  \\  0.8 \pm 1.0 \\  \times \evec \times 10^{-2},
\quad
\left[\hat  c_{Hd}\right]_{ii}   = \bvec  -6  \pm 32 \\ -7  \pm 10 \\ -4.6 \pm  1.6 \evec \times 10^{-2}, 
\eeq
with the correlation matrix given in \eref{rho21_warsaw}. 
We stress that only the combinations in~\eref{EWPT_chats} are constrained by the pole observables.
Conversely,  the pole observables calculated in the Warsaw basis are completely independent on the Wilson coefficients along the flat directions defined by $[\hat c_{Hf}]_{ij} = 0$.
Therefore,  individually, $c_{Hf}$, $c_{WB}$, and $c_T$ cannot be constrained by the pole observables alone.
To this end, the input from off-pole and/or Higgs observables has to be included. 
For example, including the LEP-2 WW production data breaks the degeneracy and allows one to separately constrain  $c_{Hf}$, $c_{WB}$, and $c_T$~\cite{Pomarol:2013zra,Falkowski:2014tna}. 

\subsection{SILH' basis}
\label{app:silhp}

Another popular choice of dimension-6 operators is  the so-called SILH basis ~\cite{Giudice:2007fh,Contino:2013kra}.
Here we discuss a variant used in  Ref.~\cite{Pomarol:2013zra} where the 4-derivative dimension-6 operators used in the original reference are absent, 
and we refer to it as the {\em SILH' basis}. 
The  Lagrangian is written as  $\cL_{\rm S'B} =  \cL^{\rm SM} + {1 \over v^2} \sum_i s_i O_i$. 
Compared to the Warsaw basis defined in \aref{warsaw},  SILH'  contains the following 6 new operators: 
\bea
 \label{eq:silhadd}
O_W &=&  \frac{ig}{2}\left(H^\dagger \sigma^i {\overleftrightarrow {D_\mu}} H \right) D_\nu W_{\mu\nu}^i,
\nnl
O_B &=&  \frac{ig'}{2}\left(H^\dagger  {\overleftrightarrow { D_\mu}} H \right) \partial_\nu B_{\mu\nu} ,
\nnl
O_{HW} &=&   i g \left(D_\mu H^\dagger \sigma^i  D_\nu H \right) W^i_{\mu\nu},
\nnl
O_{HB} &=&   ig'  \left(D_\mu H ^\dagger D_\nu H \right) B_{\mu\nu},
\nnl
O_{\widetilde{HW}} &=&   i g \left(D_\mu H^\dagger \sigma^i  D_\nu H \right) \widetilde W^i_{\mu\nu},
\nnl
O_{\widetilde{HB}} &=&   ig'  \left(D_\mu H ^\dagger D_\nu H \right)  \widetilde  B_{\mu\nu},
\eea
The remaining operators are the ones from \tref{warsaw}, with the exception of  4 bosonic operators $O_{WW}$, $O_{\widetilde{ WW}}$,  $O_{WB}$, $O_{\widetilde{WB}}$,  and 2 vertex operators $[O_{H \ell}]_{11}$, $[O_{H \ell}']_{11}$.  
This way the number of independent Wilson coefficients is the same in the Warsaw and SILH' basis. 

In the SILH's basis, the constraints from $W$- and $Z$-pole observables on the Wilson coefficients of dimension-6 operators are as follows:
\beq
 [s_{\ell\ell}]_{1221}  = (4.8 \pm 1.6) \times 10^{-2},  \quad {s_W + s_B \over 2} = -0.43 \pm 0.26, \quad s_T = (-1.03 \pm 0.63) \times 10^{-2},  
 \nonumber 
\eeq 
\beq
\left[s'_{H\ell}\right]_{ii} = \bvec  0 \\ -0.36 \pm 0.92 \\ 3.0 \pm 1.3 \evec  \times 10^{-2},
\quad
\left[s_{H\ell}\right]_{ii}  = \bvec  0  \\  0.29 \pm 0.95  \\  -3.0 \pm 1.3 \evec  \times 10^{-2},
 \nonumber 
\eeq
\beq
\left[s_{He}\right]_{ii}  = \bvec   -2.0 \pm 1.3 \\  -2.1 \pm 1.3 \\  -2.2 \pm 1.3 \evec  \times 10^{-2},
 \nonumber 
\eeq
\beq
\left[s'_{Hq}\right]_{ii}  = \bvec  1.2 \pm 2.8 \\ -0.1 \pm 2.9 \\ 0.4 \pm 3.8 \evec  \times 10^{-2},
\quad
\left[s_{Hq}\right]_{ii}  = \bvec  2.1 \pm 7.1  \\ -0.4 \pm 2.9 \\  0.3 \pm 3.8 \evec \times 10^{-2},
 \nonumber 
\eeq
\beq
\label{eq:results_silhp}
\left[s_{Hu}\right]_{ii}  = \bvec  -1 \pm 10  \\  2.2 \pm 1.3 \\  \times \evec \times 10^{-2},
\quad
\left[s_{Hd}\right]_{ii}   = \bvec  -6  \pm 32 \\ -7  \pm 10 \\ -5.3 \pm  1.7 \evec \times 10^{-2}, 
\eeq
with the correlation matrix given in \eref{rho21_silhp}.
Note that $\left[s'_{H\ell}\right]_{11}=\left[s_{H\ell}\right]_{11}=0$ by the definition of the SILH' basis.   
All the remaining vertex operators $O_{Hf}$ and $O_{Hf}'$ are separately constrained by the pole observables, unlike in the Warsaw basis. 
It is worth noting that the constraints on the  combination $s_W + s_B$ (related to the Peskin-Takeuchi S-parameter) are loose when marginalized over Wilson coefficients of other dimension-6 operators. 
However, large deviations of $s_W + s_B$ from zero have to be strongly correlated with deviations in $[s_{He}]_{ii}$ and $s_T$
The combination $s_W - s_B$ is not constrained by the pole observables at all.

\section{Correlation matrix}
\label{app:rho21}

Here we quote the various correlation matrices described in Sec.~\ref{sec:res}.
The rows and columns correspond to the order the results are presented in Eqs.~(\ref{eq:dgl}), (\ref{eq:flavorblind}), (\ref{eq:mfv}), (\ref{eq:composite}), (\ref{eq:results_warsaw}), and (\ref{eq:results_silhp}). 

\begin{landscape}
\beqa\label{eq:rho21}
&&\rho = \\
&&\left ( \begin{array}{ccc|ccc|ccc|ccc|cc|ccc|ccc}
1.   &-0.12&-0.63&-0.10&-0.03& 0.01& 0.07&-0.06&-0.04&-0.02& 0   & 0   &-0.03& 0.01& -0.02&-0.03& 0.02&-0.05&-0.03 & 0 \\
\cdot&1.   &-0.56&-0.11&-0.04& 0.01& 0.08&-0.06&-0.04&-0.02& 0   & 0   &-0.03& 0.01 & -0.02&-0.03& 0.02&-0.05&-0.04  & 0  \\
\cdot&\cdot&1.   &-0.10&-0.03& 0.01& 0.07&-0.05&-0.04& 0.01&-0.01& 0   & 0.02&-0.01& 0.01& 0.03& 0.02& 0.04& 0.03 & 0.01    \\
\hline
\cdot&\cdot&\cdot&1.   &-0.10&-0.07& 0.17&-0.05& 0.03& 0.02& 0.08&-0.02& 0.03& 0.09& 0.02& 0.03&-0.38& 0.05& 0.03&-0.37  \\
\cdot&\cdot&\cdot&\cdot&1.   & 0.07&-0.06& 0.90&-0.04& 0   &-0.02 & 0   & 0   &-0.01 & 0   & 0.01& 0.08& 0 & 0 & 0.05   \\
\cdot&\cdot&\cdot&\cdot&\cdot&1.   & 0.02&-0.03& 0.41&-0.01&-0.02  & 0   &-0.01& 0  & 0   & 0   & 0.08&-0.01&-0.01& 0.01  \\
\hline
\cdot&\cdot&\cdot&\cdot&\cdot&\cdot&1.   &-0.08&-0.04&-0.01& 0.07&-0.02&-0.01& 0.12&-0.01&-0.01&-0.36&-0.02&-0.01&-0.40  \\
\cdot&\cdot&\cdot&\cdot&\cdot&\cdot&\cdot&1.   & 0.04& 0.01& 0   & 0   & 0.01&-0.02  & 0.01& 0.01& 0.02& 0.02& 0.02& 0.05  \\
\cdot&\cdot&\cdot&\cdot&\cdot&\cdot&\cdot&\cdot&1.   & 0.01& 0.02& 0   & 0.01&-0.01& 0.01& 0.01&-0.05& 0.02& 0.02& 0.01     \\
\hline
\cdot&\cdot&\cdot&\cdot&\cdot&\cdot&\cdot&\cdot&\cdot&1.   &-0.07& 0   & 0.72& 0.06& 0.79&-0.06&-0.01& 0.76&-0.12& 0     \\
\cdot&\cdot&\cdot&\cdot&\cdot&\cdot&\cdot&\cdot&\cdot&\cdot&1.   &0& 0.03& 0.29 &-0.04& 0.10&-0.11& 0.03& 0.03&-0.15  \\
\cdot&\cdot&\cdot&\cdot&\cdot&\cdot&\cdot&\cdot&\cdot&\cdot&\cdot&1.   & 0   &-0.01& 0   & 0   & 0.04& 0   & 0   & 0.04 \\
\hline
\cdot&\cdot&\cdot&\cdot&\cdot&\cdot&\cdot&\cdot&\cdot&\cdot&\cdot&\cdot&1.   & 0.03 & 0.71&-0.21&-0.01& 0.92&-0.15 &-0.01 \\
\cdot&\cdot&\cdot&\cdot&\cdot&\cdot&\cdot&\cdot&\cdot&\cdot&\cdot&\cdot&\cdot&1.    & 0.03& 0.03&-0.19& 0.06& 0.04  &-0.15\\
\hline
\cdot&\cdot&\cdot&\cdot&\cdot&\cdot&\cdot&\cdot&\cdot&\cdot&\cdot&\cdot&\cdot&\cdot&1.   &-0.63&-0.01& 0.66& 0.01  & 0   \\
\cdot&\cdot&\cdot&\cdot&\cdot&\cdot&\cdot&\cdot&\cdot&\cdot&\cdot&\cdot&\cdot&\cdot&\cdot&1.   &-0.02&-0.04&-0.03 &-0.02\\
\cdot&\cdot&\cdot&\cdot&\cdot&\cdot&\cdot&\cdot&\cdot&\cdot&\cdot&\cdot&\cdot&\cdot&\cdot&\cdot&1.   &-0.02&-0.02 & 0.89\\
\hline
\cdot&\cdot&\cdot&\cdot&\cdot&\cdot&\cdot&\cdot&\cdot&\cdot&\cdot&\cdot&\cdot&\cdot&\cdot&\cdot&\cdot&1.   &-0.32 &-0.02\\
\cdot&\cdot&\cdot&\cdot&\cdot&\cdot&\cdot&\cdot&\cdot&\cdot&\cdot&\cdot&\cdot&\cdot&\cdot&\cdot&\cdot&\cdot&1.   &-0.01 \\
\cdot&\cdot&\cdot&\cdot&\cdot&\cdot&\cdot&\cdot&\cdot&\cdot&\cdot&\cdot&\cdot&\cdot &\cdot&\cdot&\cdot&\cdot&\cdot&1.    \\
\end{array} \right ) . \nonumber
\eeqa
\end{landscape}

\beq\label{eq:rhoUni}
\rho^{\rm UNI} = \left ( \begin{array}{ccccccccc}
 1. & -0.55 & 0.15 & 0.02 & 0.03 & 0.09 & 0.06  \\
\cdot&1. & 0.34 & 0.02 & 0.05& - 0.28 & -0.34 \\
\cdot&\cdot&1. & 0.09 & 0.07 & -0.39 & -0.38  \\
\cdot&\cdot&\cdot&1. & 0.83& 0.04 & -0.11  \\
\cdot &\cdot&\cdot&\cdot&1. & -0.13 & -0.05  \\
\cdot&\cdot&\cdot&\cdot&\cdot&1. & 0.89  \\
\cdot&\cdot&\cdot&\cdot&\cdot&\cdot&1. \\
\end{array} \right ).
\eeq

\beqa\label{eq:rhoMFV}
&&\rho^{\rm MFV} = \\
&&\left(
\begin{array}{ccccccccccccc}
 1. & -0.97 & -0.11 & 0.02 & 0.05 & -0.05 & 0. & 0. & -0.11 & 0.11 & 0.01 & -0.12 & 0.12 \\
 \cdot & 1. & -0.01 & 0. & 0. & 0. & -0.01 & 0. & 0.09 & -0.09 & -0.01 & 0.1 & -0.09 \\
 \cdot & \cdot & 1. & -0.36 & 0.36 & -0.18 & 0.08 & -0.02 & 0.08 & -0.17 & 0.11 & 0.08 & -0.18 \\
 \cdot & \cdot & \cdot & 1. & -0.19 & 0.49 & -0.05 & 0.01 & -0.03 & 0.07 & -0.04 & -0.03 & 0.08 \\
 \cdot & \cdot & \cdot & \cdot & 1. & -0.35 & 0.11 & -0.03 & 0.03 & -0.15 & 0.11 & 0.04 & -0.15 \\
 \cdot & \cdot & \cdot & \cdot & \cdot & 1. & -0.03 & 0.01 & 0. & 0.04 & -0.05 & 0.01 & 0.04 \\
 \cdot & \cdot & \cdot & \cdot & \cdot & \cdot & 1. & -0.1 & 0.52 & -0.52 & 0.43 & 0.23 & -0.27 \\
 \cdot & \cdot & \cdot & \cdot & \cdot & \cdot & \cdot & 1. & -0.05 & 0.06 & -0.04 & -0.02 & 0.04 \\
 \cdot & \cdot & \cdot & \cdot & \cdot & \cdot & \cdot & \cdot & 1. & -0.96 & 0.19 & 0.9 & -0.86 \\
 \cdot & \cdot & \cdot & \cdot & \cdot & \cdot & \cdot & \cdot & \cdot & 1. & -0.23 & -0.86 & 0.91 \\
 \cdot & \cdot & \cdot & \cdot & \cdot & \cdot & \cdot & \cdot & \cdot & \cdot & 1. & 0.36 & -0.38 \\
 \cdot & \cdot & \cdot & \cdot & \cdot & \cdot & \cdot & \cdot & \cdot & \cdot & \cdot & 1. & -0.95 \\
 \cdot & \cdot & \cdot & \cdot & \cdot & \cdot & \cdot & \cdot & \cdot & \cdot & \cdot & \cdot & 1. \\
\end{array}
\right)
\, . \nonumber
\eeqa

\beqa\label{eq:rhoPC}
\rho^{\rm VL} =
&&
\left(
\begin{array}{cccccc}
 1. & -0.95 & -0.19 & -0.01 & 0.19 & 0. \\
 \cdot & 1. & 0.17 & 0.01 & -0.17 & -0.04 \\
 \cdot & \cdot & 1. & 0. & 0.07 & 0.85 \\
 \cdot & \cdot & \cdot & 1. & 0.02 & 0. \\
 \cdot & \cdot & \cdot & \cdot & 1. & -0.13 \\
 \cdot. & \cdot & \cdot & \cdot & \cdot & 1. \\
\end{array}
\right)\, .
\eeqa

\begin{landscape}
\beqa\label{eq:rho21_warsaw}
&&\rho^{\rm Warsaw} = \\
&&\left ( \begin{array}{c|ccc|ccc|ccc|ccc|ccc|cc|ccc}
1&0.7&0.63&-0.89&-0.68&-0.57&0.88&-0.07&0.1&0.07&0.01&0.05&0&0.03&0.04&0&0.04&-0.01&0.07&0.05&-0.01\\
\hline
\cdot&1&-0.11&-0.62&-0.99&0.13&0.62&-0.01&0.07&0.07&0.01&0.03&0&0.02&0.03&0&0.03&-0.01&0.05&0.03&0  \\
\cdot&\cdot&1&-0.54&0.13&-0.93&0.55&-0.01&0.08&0.07&0.01&0.03&0&0.02&0.03&0&0.03&-0.01&0.05&0.04&0    \\
\cdot&\cdot&\cdot&1&0.64&0.54&-0.98&-0.01&0.07&0.06&0&-0.03&0&-0.02&-0.03&0&-0.02&0&-0.04&-0.03&-0.01  \\
\hline
\cdot&\cdot&\cdot&\cdot&1&-0.14&-0.64&0.09&-0.06&-0.03&0&-0.03&0&-0.02&-0.03&0&-0.02&0.01&-0.04&-0.03&-0.03  \\
\cdot&\cdot&\cdot&\cdot&\cdot&1&-0.53&0.06&0.26&-0.04&0&-0.03&0&-0.02&-0.03&0&-0.03&0&-0.05&-0.03&0.02  \\
\cdot&\cdot&\cdot&\cdot&\cdot&\cdot&1&0.07&-0.05&0.03&0.01&0.03&0&0.01&0.02&0&0.02&0&0.04&0.03&0.01\\
\hline
\cdot&\cdot&\cdot&\cdot&\cdot&\cdot&\cdot&1&0.01&0.13&0.01&0&0.01&-0.01&0&-0.03&-0.01&0.08&-0.02&-0.01&-0.33 \\
\cdot&\cdot&\cdot&\cdot&\cdot&\cdot&\cdot&\cdot&1&0.08&0.01&0.02&0&0.01&0.01&0&0.01&-0.02&0.02&0.02&0.05 \\
\cdot&\cdot&\cdot&\cdot&\cdot&\cdot&\cdot&\cdot&\cdot&1&0.01&0.01&0&0.01&0.01&0&0.01&-0.02&0.02&0.02&0.01 \\
\hline
\cdot&\cdot&\cdot&\cdot&\cdot&\cdot&\cdot&\cdot&\cdot&\cdot&1&-0.94&0&0.49&-0.91&0&0.31&-0.02&0.2&0.14&0\\
\cdot&\cdot&\cdot&\cdot&\cdot&\cdot&\cdot&\cdot&\cdot&\cdot&\cdot&1&0&-0.41&0.97&0&-0.21&-0.01&-0.04&-0.03&0\\
\cdot&\cdot&\cdot&\cdot&\cdot&\cdot&\cdot&\cdot&\cdot&\cdot&\cdot&\cdot&1&0&0&-1&0&0&0&0&0 \\
\hline
\cdot&\cdot&\cdot&\cdot&\cdot&\cdot&\cdot&\cdot&\cdot&\cdot&\cdot&\cdot&\cdot&1&-0.41&0&0.75&0.04&0.74&-0.05&0\\
\cdot&\cdot&\cdot&\cdot&\cdot&\cdot&\cdot&\cdot&\cdot&\cdot&\cdot&\cdot&\cdot&\cdot&1&0&-0.2&0.07&-0.03&-0.02&-0.03   \\
\cdot&\cdot&\cdot&\cdot&\cdot&\cdot&\cdot&\cdot&\cdot&\cdot&\cdot&\cdot&\cdot&\cdot&\cdot&1&0&-0.02&0&0&0.07\\
\hline
\cdot&\cdot&\cdot&\cdot&\cdot&\cdot&\cdot&\cdot&\cdot&\cdot&\cdot&\cdot&\cdot&\cdot&\cdot&\cdot&1&0.03&0.92&-0.15&-0.01\\
\cdot&\cdot&\cdot&\cdot&\cdot&\cdot&\cdot&\cdot&\cdot&\cdot&\cdot&\cdot&\cdot&\cdot&\cdot&\cdot&\cdot&1&0.06&0.04&-0.15\\
\hline
\cdot&\cdot&\cdot&\cdot&\cdot&\cdot&\cdot&\cdot&\cdot&\cdot&\cdot&\cdot&\cdot&\cdot&\cdot&\cdot&\cdot&\cdot&1&-0.32&-0.02 \\
\cdot&\cdot&\cdot&\cdot&\cdot&\cdot&\cdot&\cdot&\cdot&\cdot&\cdot&\cdot&\cdot&\cdot&\cdot&\cdot&\cdot&\cdot&\cdot&1&-0.01   \\
\cdot&\cdot&\cdot&\cdot&\cdot&\cdot&\cdot&\cdot&\cdot&\cdot&\cdot&\cdot&\cdot&\cdot&\cdot&\cdot&\cdot&\cdot&\cdot&\cdot&1    \\
\end{array} \right ) . \nonumber
\eeqa
\end{landscape}

\begin{landscape}
\beqa\label{eq:rho21_silhp}
&&\rho^{\rm SILH'} = \\
&&\left ( \begin{array}{c|c|c|cc|cc|ccc|ccc|ccc|cc|ccc}
1&0.69&0.68&-0.08&-0.89&0.08&0.88&0.68&0.68&0.69&-0.15&-0.11&-0.12&0.01&-0.01&-0.04&-0.01&-0.44&0.08&0.08&0.16\\
\hline
\cdot&1& 0.999 &-0.78&-0.89&0.76&0.88&0.999 &0.98&0.995&-0.22&-0.19&-0.17&-0.01&-0.05&-0.06&-0.06&-0.64&0.06&0.07&0.27  \\
\hline
\cdot&\cdot&1&-0.78&-0.89&0.76&0.88&0.999 &0.98&0.995 &-0.22&-0.19&-0.16&-0.01&-0.05&-0.05&-0.06&-0.64&0.06&0.07&0.27\\
\hline
\cdot&\cdot&\cdot&1&0.43&-0.97&-0.43&-0.78&-0.75&-0.77&0.18&0.17&0.13&0.02&0.06&0.04&0.06&0.49&-0.01&-0.03&-0.19  \\
\cdot&\cdot&\cdot&\cdot&1&-0.42&-0.99&-0.89&-0.86&-0.88&0.2&0.16&0.15&0.01&0.04&0.05&0.05&0.56&-0.06&-0.07&-0.22  \\
\hline
\cdot&\cdot&\cdot&\cdot&\cdot&1&0.42&0.76&0.78&0.75&-0.17&-0.17&-0.12&-0.02&-0.06&-0.04&-0.06&-0.49&0.01&0.03&0.22  \\
\cdot&\cdot&\cdot&\cdot&\cdot&\cdot&1&0.88&0.85&0.88&-0.19&-0.16&-0.14&-0.01&-0.04&-0.05&-0.05&-0.56&0.05&0.07&0.24\\
\hline
\cdot&\cdot&\cdot&\cdot&\cdot&\cdot&\cdot&1&0.98&0.99&-0.22&-0.20&0.17&-0.01&-0.05&-0.06&-0.06&-0.64&0.05&0.07&0.26 \\
\cdot&\cdot&\cdot&\cdot&\cdot&\cdot&\cdot&\cdot&1&0.98&-0.22&-0.19&-0.16&-0.01&-0.05&-0.05&-0.05&-0.63&0.06&0.07&0.28 \\
\cdot&\cdot&\cdot&\cdot&\cdot&\cdot&\cdot&\cdot&\cdot&1&-0.22&-0.19&-0.16&-0.01&-0.05&-0.05&-0.06&-0.64&0.06&0.07&0.27 \\
\hline
\cdot&\cdot&\cdot&\cdot&\cdot&\cdot&\cdot&\cdot&\cdot&\cdot&1&-0.85&0.04&0.48&-0.88&0.01&0.32&0.13&0.18&0.12&-0.05\\
\cdot&\cdot&\cdot&\cdot&\cdot&\cdot&\cdot&\cdot&\cdot&\cdot&\cdot&1&0.03&-0.40&0.96&0.01&-0.2&0.12&-0.05&-0.04&-0.04\\
\cdot&\cdot&\cdot&\cdot&\cdot&\cdot&\cdot&\cdot&\cdot&\cdot&\cdot&\cdot&1&0&0.01&-0.97&0.01&0.1&-0.01&-0.01&-0.04 \\
\hline
\cdot&\cdot&\cdot&\cdot&\cdot&\cdot&\cdot&\cdot&\cdot&\cdot&\cdot&\cdot&\cdot&1&-0.41&0&0.75&0.04&0.74&-0.05&0\\
\cdot&\cdot&\cdot&\cdot&\cdot&\cdot&\cdot&\cdot&\cdot&\cdot&\cdot&\cdot&\cdot&\cdot&1&0&-0.2&0.08&-0.04&-0.03&-0.05   \\
\cdot&\cdot&\cdot&\cdot&\cdot&\cdot&\cdot&\cdot&\cdot&\cdot&\cdot&\cdot&\cdot&\cdot&\cdot&1&0&0.02&0&-0.01&0.06\\
\hline
\cdot&\cdot&\cdot&\cdot&\cdot&\cdot&\cdot&\cdot&\cdot&\cdot&\cdot&\cdot&\cdot&\cdot&\cdot&\cdot&1&0.06&0.92&-0.16&-0.03\\
\cdot&\cdot&\cdot&\cdot&\cdot&\cdot&\cdot&\cdot&\cdot&\cdot&\cdot&\cdot&\cdot&\cdot&\cdot&\cdot&\cdot&1&0.01&-0.01&-0.29\\
\hline
\cdot&\cdot&\cdot&\cdot&\cdot&\cdot&\cdot&\cdot&\cdot&\cdot&\cdot&\cdot&\cdot&\cdot&\cdot&\cdot&\cdot&\cdot&1&-0.32&0 \\
\cdot&\cdot&\cdot&\cdot&\cdot&\cdot&\cdot&\cdot&\cdot&\cdot&\cdot&\cdot&\cdot&\cdot&\cdot&\cdot&\cdot&\cdot&\cdot&1&0.01   \\
\cdot&\cdot&\cdot&\cdot&\cdot&\cdot&\cdot&\cdot&\cdot&\cdot&\cdot&\cdot&\cdot&\cdot&\cdot&\cdot&\cdot&\cdot&\cdot&\cdot&1    \\
\end{array} \right ) . \nonumber
\eeqa
\end{landscape}

\section{Low Energy Constraints on off-diagonal $Z$ coupling to quarks }
\label{app:indirectZoff}
\setcounter{equation}{0}

Low energy processes, such as meson mixing or rare decays, imply strong indirect bounds on tree-level $Z$ off-off diagonal couplings. Assuming alignment, these arise only in the left handed currents. We thus consider only $[\delta Z^{Zu}_L]_{ij}$ and $[\delta Z^{Zd}_L]_{ij}$. For simplicity, we assume these parameters to be real.

For the up sector, the strongest bound is arising from charm-mixing, we follow~\cite{Gedalia:2009kh} (and the recent results in Eqs.~(62)--(63) of~\cite{Azatov:2014lha}) and find that
\begin{align} \label{eq:ZoffIndirectUp}
&	[\delta g^{Zu}_L]_{12}\lesssim 8.4\times 10^{-5}\, , \
\end{align}
where the NP is allowed to saturate the 1$\sigma$ bound on the mixing parameters.  For the down sector, following~\cite{Buras:2012jb}, the strongest constrained are coming from $\Delta M_K= (0.5392\pm 0.0009) \times 10^{-2}$pb$^{-1}$~\cite{Beringer:1900zz},  ${\rm Br}(B_d \to \mu^+ \mu^-)<6.3 \times 10^{-10}$ and  ${\rm Br}(B_s \to \mu^+ \mu^-)=(3.1\pm0.7) \times 10^{-9}$~\cite{Amhis:2014hma}. We find that
\begin{align} \label{eq:ZoffIndirectDw}
	[\delta g^{Zd}_L]_{12}\lesssim 1.4\times 10^{-4}\, , \quad
&	[\delta g^{Zd}_L]_{13}\lesssim 1.5\times 10^{-4}\, , \quad
	[\delta g^{Zd}_L]_{23}\lesssim 4.6\times 10^{-4}\, , \ 		
\end{align}
is allowed at 95\%~CL.

\bibliographystyle{JHEP}
\bibliography{fnupaper_arxiv_v2}

\end{document}